\documentclass[12pt]{article}
\usepackage[margin=1in]{geometry}

\usepackage[utf8]{inputenc}
\usepackage[algo2e,norelsize]{algorithm2e} 
\SetAlCapNameFnt{\scriptsize}
\SetAlCapFnt{\scriptsize}
\usepackage{textgreek}
\usepackage{textcomp}
\usepackage{amssymb}
\usepackage{caption}
\usepackage{subcaption}
\usepackage{amsmath}
\DeclareMathOperator*{\argmax}{argmax}
\DeclareMathOperator*{\argmin}{argmin}
\usepackage{color}
\usepackage{footnote}
\usepackage{algorithm}
\usepackage{multirow} 
\usepackage{booktabs}
\usepackage{graphicx}
\usepackage{amssymb}
\usepackage{amsbsy}
\usepackage{array}
\usepackage{longtable}
\usepackage{epstopdf}
\usepackage{pbox}
\usepackage{breqn}
\usepackage{mathrsfs}
\usepackage{multicol}
\usepackage{supertabular}
\usepackage{enumerate}
\usepackage{enumitem}
\usepackage{bbm}
\usepackage{url}
\usepackage[justification=centering]{caption}
\usepackage{placeins}
\usepackage{dsfont}
\usepackage{natbib}

\newtheorem{defn}{Definition}
\newtheorem{thm}{Theorem}

\newtheorem{prop}{Proposition}

\title{\LARGE \bf Cost-aware Defense for Parallel Server Systems against Reliability and Security Failures} 

\author{Qian Xie, Jiayi Wang and Li Jin
\thanks{This work was in part supported by US NSF Award CMMI-1949710, C2SMART University Transportation Center, NYU Tandon School of Engineering, SJTU-UM Joint Institute, J. Wu \& J. Sun Endowment Fund, and Cornell University McMullen Fellowship.   
Q. Xie is with the School of Operations Research and Information Engineering, Cornell University, USA.
J. Wang is with the Department of Electrical and Computer Engineering, University of California, San Diego, USA.
L. Jin is with the UM Joint Institute, and the Department of Automation, Shanghai Jiao Tong University, China.
Q. Xie and L. Jin were with the Tandon School of Engineering, New York University, USA.
J. Wang was with the UM Joint Institute, Shanghai Jiao Tong University, China.
(emails: qx66@cornell.edu, jiw139@ucsd.edu, li.jin@sjtu.edu.cn).}
}

\begin{document}
\maketitle
\thispagestyle{plain}
\pagestyle{plain}

\begin{abstract}                          
Parallel server systems in transportation, manufacturing, and computing heavily rely on dynamic routing using connected cyber components for computation and communication. Yet, these components remain vulnerable to random malfunctions and malicious attacks, motivating the need for fault-tolerant dynamic routing that are both traffic-stabilizing and cost-efficient.
In this paper, we consider a parallel server system with dynamic routing subject to reliability and stability failures. For the reliability setting, we consider an infinite-horizon Markov decision process where the system operator strategically activates protection mechanism upon each job arrival based on traffic state observations. We prove an optimal deterministic threshold protecting policy exists based on dynamic programming recursion of the HJB equation. For the security setting, we extend the model to an infinite-horizon stochastic game where the attacker strategically manipulates routing assignment. We show that both players follow a threshold strategy at every Markov perfect equilibrium. For both failure settings, we also analyze the stability of the traffic queues under control. Finally, we develop approximate dynamic programming algorithms to compute the optimal/equilibrium policies, supplemented with numerical examples and experiments for validation and illustration.
\end{abstract}



\section{Introduction}

Parallel server system is a typical model characterizing service systems of multiple servers, each with a waiting queue. Real-world instances include packet switching networks \citep{neely2001packet,gupta2007analysis}, manufacturing systems \citep{iravani2011capability}, transportation facilities \citep{jin2018stability}, etc.
These systems use feedback from state observations to generate routing decisions that ensure stability and improve throughput. Such feedback routing heavily relies on connected cyber components for data collection and transmission. However, these cyber components face persistent threats from malfunctions and manipulations \citep{cardenas2009challenges}.

Malfunctions can arise from technical issues like network congestion, server unresponsiveness, packet loss, firewall restriction, signal interference, and authentication errors \citep{alpcan2010network}, or malicious attacks like Denial-of-Service (DoS) \citep{wang2007queueing,al2012survey} that overwhelm servers with excessive traffic and cut off state observations. In such scenarios, the system operator may fail to deliver routing instructions. To illustrate the cause and impact of routing malfunctions, consider two real-world motivating examples:
\begin{enumerate}[noitemsep,topsep=0pt]
\item \textbf{Transportation:} Imagine a vehicle experiencing a failure in receiving routing information from a navigation app due to network connection issues. In this situation, drivers often resort to independent routing decisions based on personal preferences, such as route types, tolls, scenery, and familiarity.
\item \textbf{Manufacturing:} Similarly, in a production line, where production units are supposed to be routed to the shortest queue based on real-time routing information, a communication failure or breakdown can trigger a fallback mechanism \citep{fraile2018trustworthy}, leading to a random assignment to a default queue. 
\end{enumerate}
The two examples represent two potential outcomes\footnotemark: (i) initiating a routing decision based on individual preferences, historical data, or random selection; (ii) joining a default queue predetermined by a fallback mechanism. Notably, from the system perspective, the routing choices in the former outcome exhibit a random nature.
\footnotetext{Our model does not apply to failure scenarios in which arrivals experience delays or are abandoned, say packet loss in computer networks.}

Manipulations, on the other hand, describe strategic attacks from adversaries with selfish or malicious intent. These include (i) \emph{spoofing attacks} that directly send deceptive routing instructions to arrivals by impersonating the system operator and (ii) \emph{falsification attacks} that inject misleading queue length data or create fictitious traffic, indirectly influencing the system operator's routing decisions \citep{feng2022cybersecurity,al2012survey,sakiz2017survey}.
For instance, a simulated traffic jam can cause motorists to deviate from their planned routes \citep{lidar2014technion}. Transportation infrastructure information (e.g., traffic sensors, traffic lights) and vehicle communications can also be intruded and manipulated \citep{feng2022cybersecurity,sakiz2017survey,al2012survey}.
Similar security risks also exist in industrial control \citep{barrere2020measuring,fraile2018trustworthy} and communication systems \citep{alpcan2010network,manshaei2013game,de2015input}. 


Real-world service systems facing failures will not be accepted by authorities, industry, and the public, unless security issues are well addressed.
However, cyber security risks have not been sufficiently studied in conjunction with the physical queuing dynamics.
Furthermore, perfectly avoid cyber failures is economically infeasible and technically unnecessary. Therefore, it is crucial to understand the impact due to such threats and to design practical defense mechanisms.
In practice, these defense mechanisms can be implemented with dynamic activation/deactivation of prevention/detection measures such as robust data validation, routing instruction encryption, and strict security protocol adherence \citep{cardenas2009challenges,manshaei2013game}. Nonetheless, these actions, while active, entail ongoing technological costs on computational resources, network bandwidth, energy consumption, and maintenance efforts, etc.

In response to such concerns, we try to address the following two research questions:
\begin{enumerate}[label=(\roman*)]
    \item \emph{How to model the security vulnerabilities and quantify the security risks for parallel queuing systems?}
    \item \emph{How to design traffic-stabilizing, cost-efficient defense strategies against failures?}
\end{enumerate}
\vspace{-0.2cm}
For the first question, we consider two scenarios of failures, viz. \emph{reliability failures} and \emph{security failures}. We formulate the security risks in terms of failure-induced queuing delays and defending costs.
For the second question, we analyze the stability criteria of the failure-prone system with defense, and characterize the structure of the cost-efficient strategies.
We also develop algorithms to compute such strategies, and discuss how to incorporate the stability condition.
Our results are demonstrated via a series of numerical examples and simulations. 



This paper is related to two lines of work: queuing control and game theory.
On the queuing side, the majority of the existing analysis and design are based on perfect observation of the states (i.e., queue lengths) and perfect implementation of the control \citep{ephremides1980simple,halfin1985shortest,eschenfeldt2018join,gupta2007analysis,knessl1986two}.
Besides, researchers have noted the impact of delayed \citep{kuri1995optimal,mehdian2017join}, erroneous \citep{beutler1989routing}, or decentralized information \citep{ouyang2015signaling}.
Although these results provide hints for our problem, they do not directly apply to the security setting with failures such as imperfect sensing (state observation) and imperfect control implementation.
On the game side, a variety of game-theoretic models have been applied to studying cyber-physical security in transportation \citep{tang2020security,laszka2019detection} and communication \citep{bohacek2007game,alpcan2010network,manshaei2013game}. However, to the best of our knowledge, security risks of queuing systems have not been well studied from a combined game-theoretic and queuing-control perspective, which is essential for capturing the coupling between the queuing dynamics and the attacker-defender interactions.

\begin{figure}[htbp]
\centering
\begin{subfigure}{0.65\textwidth}
    \includegraphics[width=\textwidth]{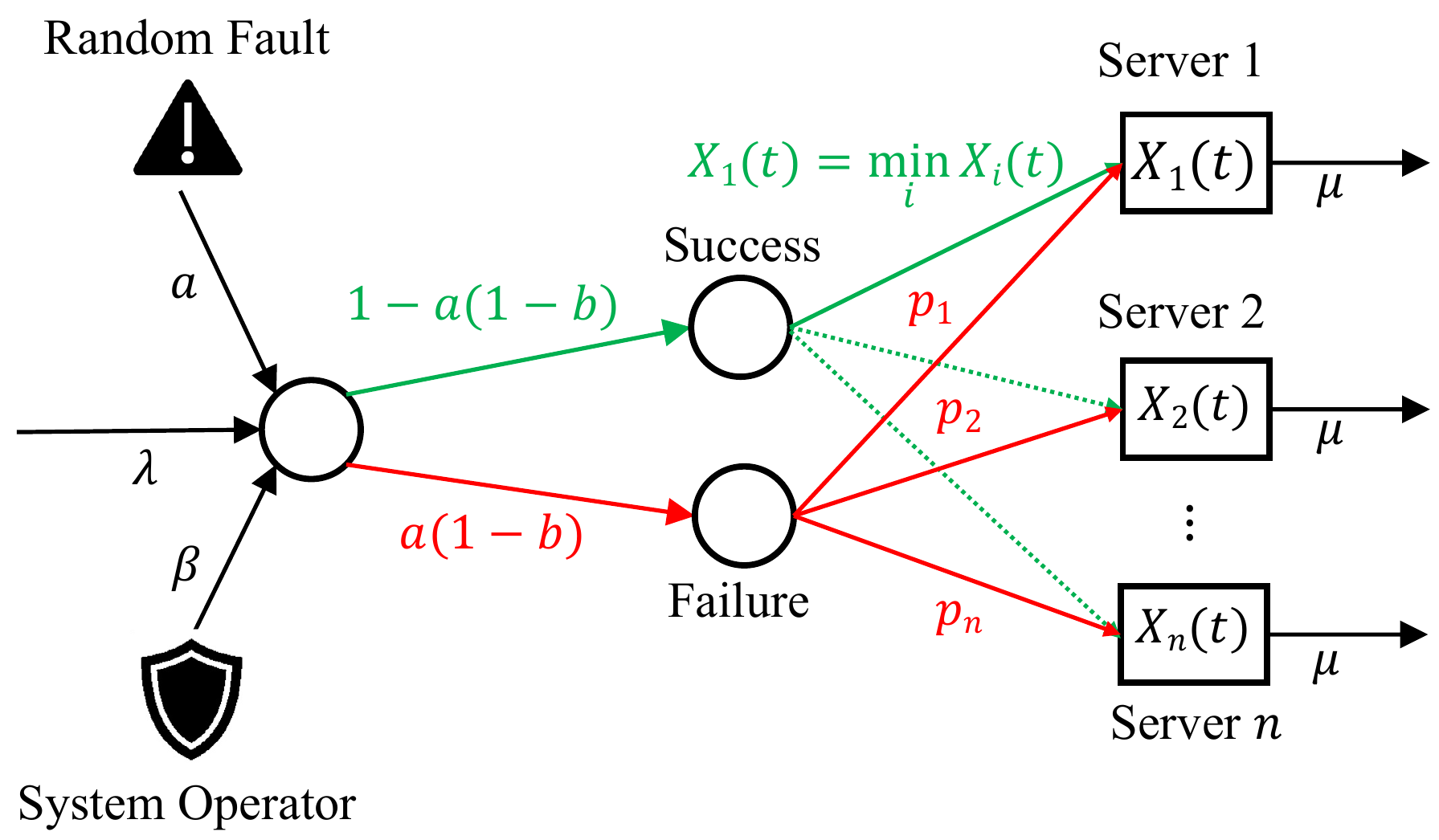}
    \caption{Scenario of reliability failures.}
    \label{fig:reliability}
\end{subfigure}

\begin{subfigure}{0.65\textwidth}
    \includegraphics[width=\textwidth]{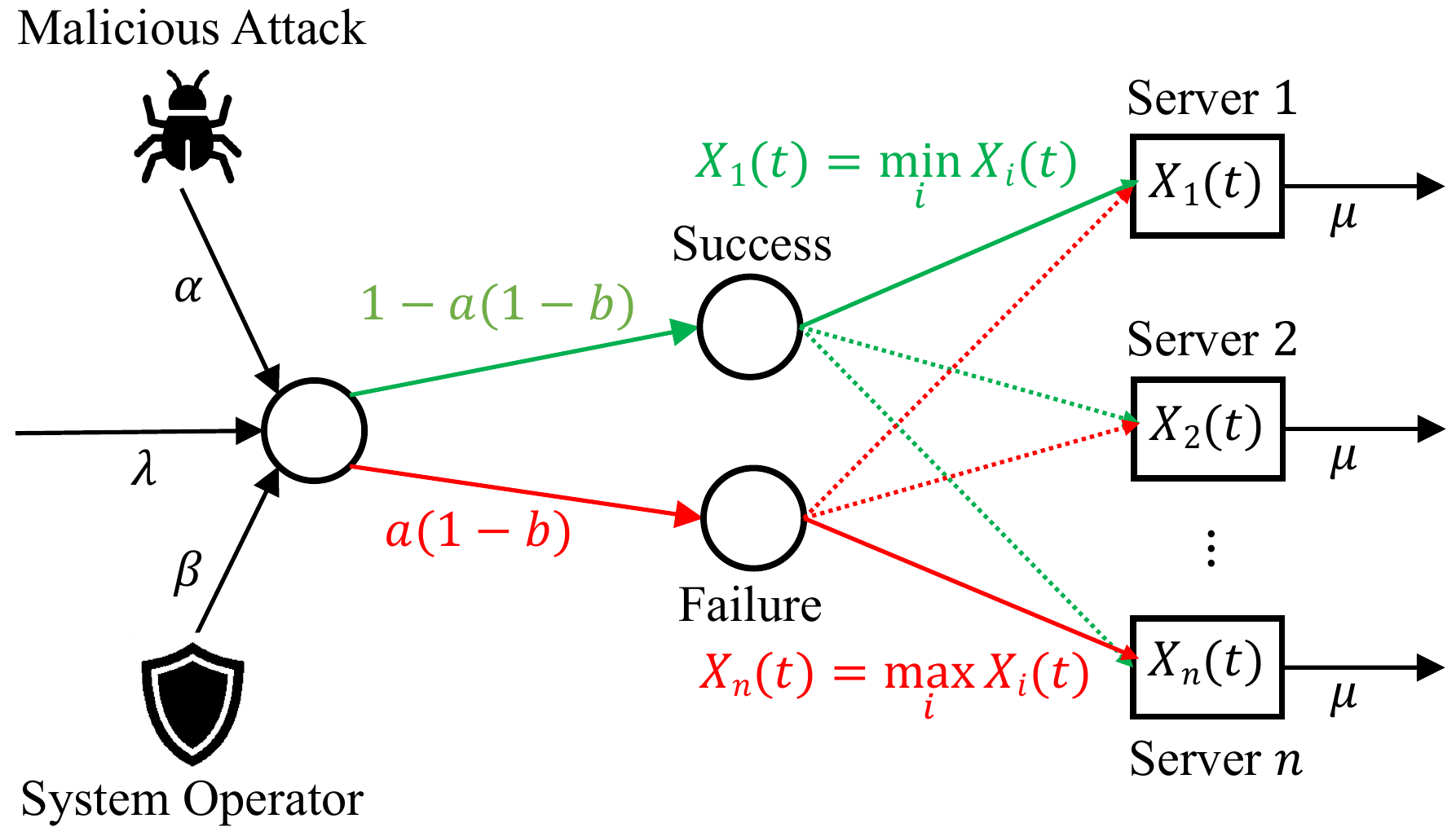}
    \caption{Scenario of security failures.}
    \label{fig:security}
\end{subfigure}
\caption{An $n$-queue system with shortest-queue routing under reliability/security failures. See Section~\ref{sec:model} for definitions of notations.}
\end{figure}

Our model includes two parts: the physical model (parallel server system) and the cyber model (dynamic routing\footnotemark subject to failures).
Specifically, we investigate the following two failure scenarios:
\begin{enumerate}
    \item \textbf{Reliability failures.} A fault may occur each routing with a constant probability and the system operator can choose to activate protection for each arrival. In the event of a routing malfunction and the absence of activated protection, the arrival joins a random queue following certain probabilities; see Fig.~\ref{fig:reliability}.
    \item \textbf{Security failures.} An adversary can launch an attack to each routing using a feedback strategy and the system operator can choose to activate defense for each arrival. In the event of an effective attack and the absence of activated defense, the arrival joins an adversary-desired queue, with the worst-case scenario being the longest queue; see Fig.~\ref{fig:security}.
\end{enumerate}
\footnotetext{Dynamic routing is a classical feedback control strategy that assigns jobs to one of the parallel queues according to the current system state (queue length).}
The system operator can protect/defend\footnotemark the routing instructions for incoming jobs based on the state observation.
\footnotetext{In the rest of the paper, we use the word ``protect'' for the reliability setting and ``defend'' for the security setting.}
The activation of protection/defense mechanisms induces a technological cost rate on the operator.
For the reliability setting, we formulate the operator's trade-off between queuing costs and protection costs as an infinite-horizon, continuous-time, and discrete-state \emph{Markov decision process}.
For the security setting, we formulate the interaction between the attacker and the operator as an infinite-horizon \emph{stochastic game}.

To study the stability of the queuing system, previous works typically relied on characterization or approximation of the steady-state distribution of queuing states \citep{foley2001join}. However, this approach is hard to integrate with failure models.
Additionally, the steady-state distribution of queuing systems with state-dependent transition rates is intricate.
In response to these challenges, we adopt a \emph{Lyapunov function}-based approach which has been applied to queuing systems in no-failure scenarios \citep{kumar1995stability,dai1995stability,eryilmaz2007fair,xie2022stabilizing}
and enables us to derive \emph{stability criterion} for queuing systems under control and to obtain upper bounds for the mean number of jobs in the system \citep{meyn1993stability}.

To analyze the cost efficiency of the system operator's decision, we formulate the optimization problem in terms of queuing and protection costs and then derive its \emph{Hamiltonian-Jacobian-Bellman (HJB) equation} \citep{yong1999stochastic}. 
We show that the optimal protecting policy under reliability failures is a deterministic \emph{threshold policy}: the operator either protects or does not protect, according to threshold functions in the multi-dimensional state space.
Similar approaches have been discussed in \citep{bertsekas1999dynamic,hajek1984optimal} for two queues, and we generalize the analysis to $n$ queues and to failure-prone settings. 
For the attacker-defender game, we again use HJB equation to show the threshold properties of the \emph{Markov perfect equilibria}.

The above analysis leads to useful insights for designing strategic protection/defense that are both stabilizing and cost-efficient. A key finding is that the system operator has a higher incentive to protect/defend if the queues are more ``imbalanced''.
In addition, our numerical analysis shows that 1) the incentive to protect increases with the failure probability, decreases with the technological cost, and increases with the utilization ratio; 2) the optimal protecting policy performs better than static policies such as never protect and always protect.
We also note that the optimal decision is not always stabilizing.
Considering this, we propose how to compute the stability-constrained optimal policy by imposing the stability condition on the HJB equation.

\textbf{Our contributions} lie in the following three aspects:
\begin{itemize}
    \item \textbf{Modeling:} 1) We build a framework for modeling the cyber-physical vulnerabilities of queuing systems with feedback control (dynamic routing) subject to reliability/security failures. 2) We propose a formulation of protection under reliability failures as an infinite-horizon Markov decision process and defense against security failures as an attacker-defender stochastic game.
    
    \item \textbf{Analysis:} 1) We provide stability criteria under failures and control based on Lyapunov functions. 2) We show the threshold properties of the optimal protection and the game equilibria on multidimensional state space based on HJB equations.
    
    \item \textbf{Design:} Our theoretical results provide insights on the design of traffic-stabilizing and cost-efficient protecting policy and defending response. We also propose approximate dynamic programming algorithms to numerically compute the optimal policy and equilibrium strategies.
\end{itemize}

The rest of this paper is organized as follows.
Section~\ref{sec:model} introduces the queuing model and the failure models.
Section~\ref{sec:reliability} studies protection against reliability failures and Section~\ref{sec:security} analyzes defense against security failures.
Section~\ref{sec:conclude} gives a concluding remark.
\section{Parallel servers and failure models}\label{sec:model}

\subsection{Parallel server system}
Consider a queuing system with $n$ identical servers in parallel. Jobs (e.g., vehicles, customers, production units) arrive according to a Poisson process of rate $\lambda>0$. Each server serves jobs at an exponential rate of $\mu>0$.
We use $X(t)=\begin{bmatrix}X_1(t) & X_2(t) & \cdots & X_n(t)\end{bmatrix}^T$ to denote the number of jobs at time $t$, either waiting or being served, in the $n$ servers, respectively.
The state space of the parallel queuing system is $\mathbb Z_{\ge0}^n$. Specifically, the initial system state (queue length) is $X(0)=x=[x_1 \ x_2 \ \cdots x_n]^T$. 

We use $x+(-)e_i$ to denote adding 1 to (subtracting 1 from) $x_i$. Since the queue lengths are always non-negative, i.e. $x_i\geq0$, we use $(x-e_i)^+=\max(x-e_i,0)$ to avoid the case that subtracting 1 makes the element negative. 
Let $x_{\min}=\min_ix_i$ and $x_{\max}=\max_ix_i$. 
We use $x_{-i}$ to denote variables in $x$ other than $x_i$, and we use the notation $x+e_{\min}$ when adding $1$ to $x_{\min}$ while keeping $x_{-i}$ the same.
We call $x$ a diagonal vector if $x_1=x_2=\cdots=x_n$ and a non-diagonal vector otherwise. Denote the one-norm of the vector $x$ as $||x||_1:=x_1+x_2+\cdots+x_n$. Then $||X(t)||_1$ means the total number of jobs in the system at time $t$. We use $x\succ\boldsymbol{0}$ to denote that $x$ is not a zero vector, i.e., $||x||_1>0$.

Without any failures, any incoming job is allocated to the shortest queue. If there are multiple shortest queues, then the job is randomly allocated to one of them with (not necessarily equal) probabilities.

\subsection{Reliability failures}
Suppose that upon the arrival of a job to the system, a fault may occur with a constant probability denoted by $a\in(0,1]$. 
In the absence of any protection mechanism, the fault would lead to a faulty routing instruction. Consequently, the job joins a random queue with respective probabilities\footnotemark $p_1, p_2, \cdots, p_n\in[0,1]$, where $\sum_{i=1}^np_i=1$. For convenience, we define 
$$p_{\max}:=\max(p_1, p_2, \cdots, p_n).$$
The system operator can decide whether to protect an arriving job to ensure its optimal routing, i.e., the shortest-queue routing, as illustrated in Fig.~\ref{fig:reliability}. However, such protection comes at the cost of a rate $c_b>0$.
\footnotetext{The system operator is assumed to know the values of random probabilities $p$, which can be estimated using the historical data and statistical techniques, or predetermined by a fallback mechanism, contingent upon the specific contexts.}

In this scenario, the system operator faces a trade-off between the queuing cost and the protection cost. We formulate this problem as an \emph{infinite-horizon continuous-time Markov decision process} with the queue lengths as the states.
The system operator adopts a \emph{Markovian policy} denoted by $\beta:\mathbb Z_{\ge0}^n\to\Delta(\{\text{NP},\text{P}\})$ where $\Delta(\{\text{NP},\text{P}\}):=\{(1-b, b): b\in[0,1]\}$ represents the probability distribution over the action set \{not protect, protect\}.
The state-dependent protecting probability at state $x\in\mathbb Z_{\ge0}^n$ is denoted as $b(x):=\beta(\text{P}\mid x)$. For simplicity, when the policy is deterministic, we write the mapping as $\beta:\mathbb Z_{\ge0}^n\to\{\text{NP},\text{P}\}$.
The state-transition matrix $P_{\mathrm{R}}:\mathbb Z_{\ge0}^n\times\{\mathrm{NP},\mathrm{P}\}\mapsto\Delta\left(\mathbb Z_{\ge0}^n\right)$ that captures the queuing dynamics under the protection against reliability failures is given by $P_{\mathrm{R}}(x+e_{\min}\mid x,\mathsf{b})=\left(1-a\mathbbm{1}\{\mathsf{b}=\text{NP}\}\right)\lambda$, $P_{\mathrm{R}}(x+e_{i}\mid x,\mathsf{b})=a\mathbbm{1}\{\mathsf{b}\neq\text{NP}\}p_i\lambda$, and $P_{\mathrm{R}}((x-e_i)^+\mid x,\cdot)=\mu$, $\forall i\in[n]$, where $\mathsf{b}$ is the action chosen by the system operator at state $x$.

Formally, the objective of the system operator is to find an optimal policy $\beta$ that minimizes the expected cumulative discounted cost $J(x;\beta)$ given initial state $X(0)=x$:
\begin{align}
    J^*(x)
    :=&\min_{\beta}J(x;\beta) \nonumber\\
    =\min_{\beta}&\ \mathbb{E}\Big[\int_0^\infty e^{-\gamma t}C(X(t),\ B(t))dt\Big\vert X(0)=x, B(t)\nonumber\\
    \sim\beta&(X(t)),\ X(t+dt)\sim P_{\mathrm{R}}(\cdot\mid X(t),B(t))\Big],
    \label{eq_J*}
\end{align}
where $\gamma\in(0,1)$ is the discounted factor, $B(t)\in\{\mathrm{NP},\mathrm{P}\}$ denotes the action chosen by the operator at time $t$, and $C:\mathbb Z_{\ge0}^n\times\{\mathrm{NP},\mathrm{P}\}\to\mathbb{R}_{\geq0}$ is the net cost rate defined as
$$C(\xi, \mathsf{b}):=||\xi||_1+c_b \mathbbm{1}\{\mathsf{b}=\text{P}\}.$$
\begin{defn}[Optimal protecting policy]
\label{defn:optimal}
The optimal protecting policy $\beta^*$ against reliability failures is defined as:
$$\beta^*(x):=\argmin_{\beta}J(x;\beta), \quad\forall x\in\mathbb{Z}_{\geq0}^n.$$
\end{defn}
\subsection{Security failures}
Suppose that when each job arrives, a malicious attacker is able to manipulate the routing such that the job is allocated to a non-shortest queue. For ease of presentation, we consider the attacker's best action (and thus the operator's worst case); i.e., the job goes to the longest queue, as shown in Fig.~\ref{fig:security}.
Let $\Delta(\{\text{NA},\text{A}\}):=\{(1-a, a): a\in[0,1]\}$ denote the probability distribution over the action set \{not attack, attack\}. The attacker selects a (possibly mixed) \emph{Markov strategy} $\alpha:\mathbb Z_{\ge0}^n\to\Delta(\{\text{NA},\text{A}\})$. With a slight abuse of notation, we write $a(x):=\alpha(\text{A}\mid x)$ as the state-dependent attacking probability. Note that here $a(x)$ has a different meaning from the constant fault probability $a$ in the reliability failure setting. The technological cost rate of attacking a job is $c_a>0$.

The system operator's action is similar to that in the reliability setting. The only difference is that in the security setting, the system operator knows there is a strategic attacker making decisions simultaneously.
We formulate the interaction between the attacker and the operator (also called defender) as an \emph{infinite-horizon stochastic game} with Markov strategies that do not depend on the history of states and actions.
The attacker aims to maximize the expected cumulative discounted reward $V(x;\alpha,\beta)$ given the operator's Markov strategy $\beta$:
\begin{align*}
    V_A^*(x;\beta)
    :=\max_{\alpha}&V(x;\alpha,\beta)]\nonumber\\
    =\max_{\alpha}&\mathbb{E}\Big[\int_0^\infty e^{-\gamma t}R(X(t), A(t), B(t))dt\Big|X(0)=x, A(t)\sim\alpha(X(t)), \nonumber\\
    & B(t)\sim\beta(X(t)), X(t+dt)\sim P_{\mathrm{S}}(\cdot\mid X(t),A(t),B(t))\Big],
\end{align*}
where $A(t)\in\{\mathrm{NA},\mathrm{A}\}$ and $B(t)\in\{\mathrm{NP},\mathrm{P}\}$ denote the actions chosen by the attacker and the operator respectively at time $t$, $P_{\mathrm{S}}:\mathbb Z_{\ge0}^n\times\{\mathrm{NA},\mathrm{A}\}\times\{\mathrm{NP},\mathrm{P}\}\mapsto\Delta\left(\mathbb Z_{\ge0}^n\right)$ represents the transition matrix that captures the queuing dynamics under security failures, as defined by $P_{\mathrm{S}}(\xi+e_{\max}\mid\xi,\mathsf{a}, \mathsf{b})=\mathbbm{1}\{\mathsf{a}=\text{A}\}\mathbbm{1}\{\mathsf{b}=\text{NP}\}\lambda$, $P_{\mathrm{S}}(\xi+e_{\min}\mid \xi,\mathsf{a}, \mathsf{b})=\mathbbm{1}\{\mathsf{a}\neq\text{A}\}\mathbbm{1}\{\mathsf{b}\neq\text{NP}\}\lambda$, $P_{\mathrm{S}}((\xi-e_i)^+\mid \xi,\cdot,\cdot)=\mu$, $\forall i\in[n]$, and $R:\mathbb Z_{\ge0}^n\times\{\mathrm{NA},\mathrm{A}\}\times\{\mathrm{NP},\mathrm{P}\}\to\mathbb R$ is the net reward rate defined as 
$$R(\xi, \mathsf{a}, \mathsf{b}):=||\xi||_1+c_b\mathbbm{1}\{\mathsf{b}=\text{P}\}-c_a\mathbbm{1}\{\mathsf{a}=\text{A}\}.$$
Here we model the attacker-defender game as a zero-sum game, which aligns with established security game literature \citep{alpcan2010network}. The attacker's reward comprises queuing attacking costs, along with a deduction for defending costs. This is motivated by the attacker potential interest in maximizing the operator's total operating cost, akin to competitive motives in business contests.
Similarly, the operator aims to minimize the expected cumulative discounted loss given the attacker's Markov strategy $\alpha$:
\begin{align*}
    V_B^*(x;\alpha):=\min_{\beta}V(x;\alpha,\beta).
\end{align*}
We can also define the Markov perfect equilibrium of such an attacker-defender game:
\begin{defn}[Markov perfect equilibrium]
The equilibrium attacking (resp. defending) strategy $\alpha^*$ (resp. $\beta^*$) satisfies that for each state $x\in\mathbb Z_{\ge0}^n$,
\begin{align*}
    &\alpha^*(x)=\argmax_{\alpha}V(x;\alpha,\beta^*)=\argmax_{\alpha}V_A^*(x;\beta^*),\\
    &\beta^*(x)=\argmin_{\beta}V(x;\alpha^*,\beta)=\argmin_{\beta}V_B^*(x;\alpha^*).
\end{align*}
The equilibrium value of the attacker (defender) is $V_A^*(x;\beta^*)$ (resp. $V_B^*(x;\alpha^*)$).
In particular, $(\alpha^*,\beta^*)$ is a Markov perfect equilibrium.
\end{defn}
\section{Protection against reliability failures}\label{sec:reliability}
In this section, we consider the design of the system operator's state-dependent protecting policy from two aspects: stability and optimality.

It is well known that a parallel $n$-server system is stabilizable if and only if the demand is less than the total capacity, i.e., $\lambda<n\mu$. In the following results, we will see that even this condition is met, in the absence of defense, reliability failures can still destabilize the queuing system, especially when the probability of failures is high and when the random faulty routing is highly heterogeneous; the following summarizes the above insights.

\begin{prop}\label{prp_stability}
The unprotected $n$-server system with faulty probability $a$ is stable if and only if
\begin{subequations}
\begin{align}
    \lambda<n\mu,\label{eq_unprotected1} \\
    ap_{max}\lambda<\mu.\label{eq_unprotected2}
\end{align}
\end{subequations}
Furthermore, when the system is stable, the long-time average number of jobs is upper-bounded by
$$
    \bar X:=\limsup\limits_{t\to\infty}\frac{1}{t}\int\limits_{\tau=0}^{t}\mathbb{E}[X(\tau)]d\tau\le 
    \frac{\lambda+n\mu}{2\left(\mu-\max(ap_{\max},\frac{1}{n})\lambda\right)}.
$$
\end{prop}

The next result provides a stability criterion for an $n$-server system with a given protecting policy. The proof of this result is presented in Section~\ref{sec:stability}.

\begin{thm}[Stability under reliability failures]\label{thm_stability}
Consider an $n$-server system with reliability failure probability $a>0$. Suppose the operator selects a Markovian policy $\beta:\mathbb Z_{\ge0}^n\to\Delta(\{\mathrm{NP},\mathrm{P}\})$ with protection probability $b(x):=\beta(\mathrm{P}\mid x)\in[0,1]$ at state $x\in\mathbb Z_{\ge0}^n$.
Then we have the following:
\begin{enumerate}[label=(\roman*)]
    \item The system is stable if for every non-diagonal vector $x$, the protecting probability $b(x)$ satisfies
    \begin{align}
        b(x)>1-\frac{\mu||x||_1-\lambda x_{\min}}{a\lambda\left(\sum\limits_{i=1}^np_ix_i-x_{\min}\right)}.
        \label{eq_betax}
    \end{align}
    \item When \eqref{eq_unprotected1} holds, there must exist a policy satisfying \eqref{eq_betax}.
    When \eqref{eq_unprotected1}-\eqref{eq_unprotected2} hold, every policy satisfies \eqref{eq_betax}.
    \item If \eqref{eq_betax} holds, the long-time average number of jobs in the system is upper-bounded by
    \begin{align}
        \bar X\le\frac{\lambda+n\mu}{2c},
        \label{eq_Xbar_reliability}
    \end{align}
\end{enumerate}
where 
$$c=\min\limits_{x\succ\boldsymbol{0}}\Big\{\mu-\lambda \frac{x_{\min}}{||x||_1}-a(1-b(x))\lambda\frac{\sum_{i=1}^np_ix_i-x_{\min}}{||x||_1}\Big\}.$$
\end{thm}
The next result implies a key finding: protection should be activated when queue lengths are more ``imbalanced''.

\begin{thm}[Optimal protecting policy]\label{thm:monotone}
Consider an $n$-server system subject to reliability failures. An optimal deterministic protecting policy $\beta^*$ exists. This deterministic policy is also a threshold policy characterized by $n$ threshold functions $f_m$ ($m=1,2,\cdots,n$) via
    $$b^*(x)=\mathbbm{1}\left\{\bigwedge\limits_{m=1}^n (f_m(x)>0)\right\},$$
    where for each $m=1,2,\cdots,n$,
    \begin{enumerate}[label=(\roman*)]
        \item threshold function $f_m: \mathbb Z_{\ge0}^n \to \mathbb{R}$ separates the polyhedron $\mathscr{X}_m = \{x\in\mathbb Z_{\ge0}^n\mid x_i\geq x_m,\ \forall 1\leq i\leq n\}$ into two subsets: $\{x\in\mathscr{X}_m \mid \beta^*(x)=\text{NP}\}$ and $\{x\in\mathscr{X}_m \mid \beta^*(x)=\text{P}\}$ by means of
        $$b^*(x)=\mathbbm{1}\{f_m(x)>0\}, \quad \forall x\in\mathscr{X}_m;$$
        \item in the polyhedron $\mathscr{X}_m = \{x\in\mathbb Z_{\ge0}^n\mid x_i\geq x_m,\ \forall 1\leq i\leq n\}$, the optimal protecting probability $b^*(x)$ is monotonically non-decreasing (resp. non-increasing) in $x_i$ ($\forall i\neq m$) (resp. $x_m$) while other variables $x_{-i}$ (resp. $x_{-m}$) are fixed.
    \end{enumerate}
\end{thm}

\begin{figure}[htbp]
    \centering
    \includegraphics[width=0.4\textwidth]{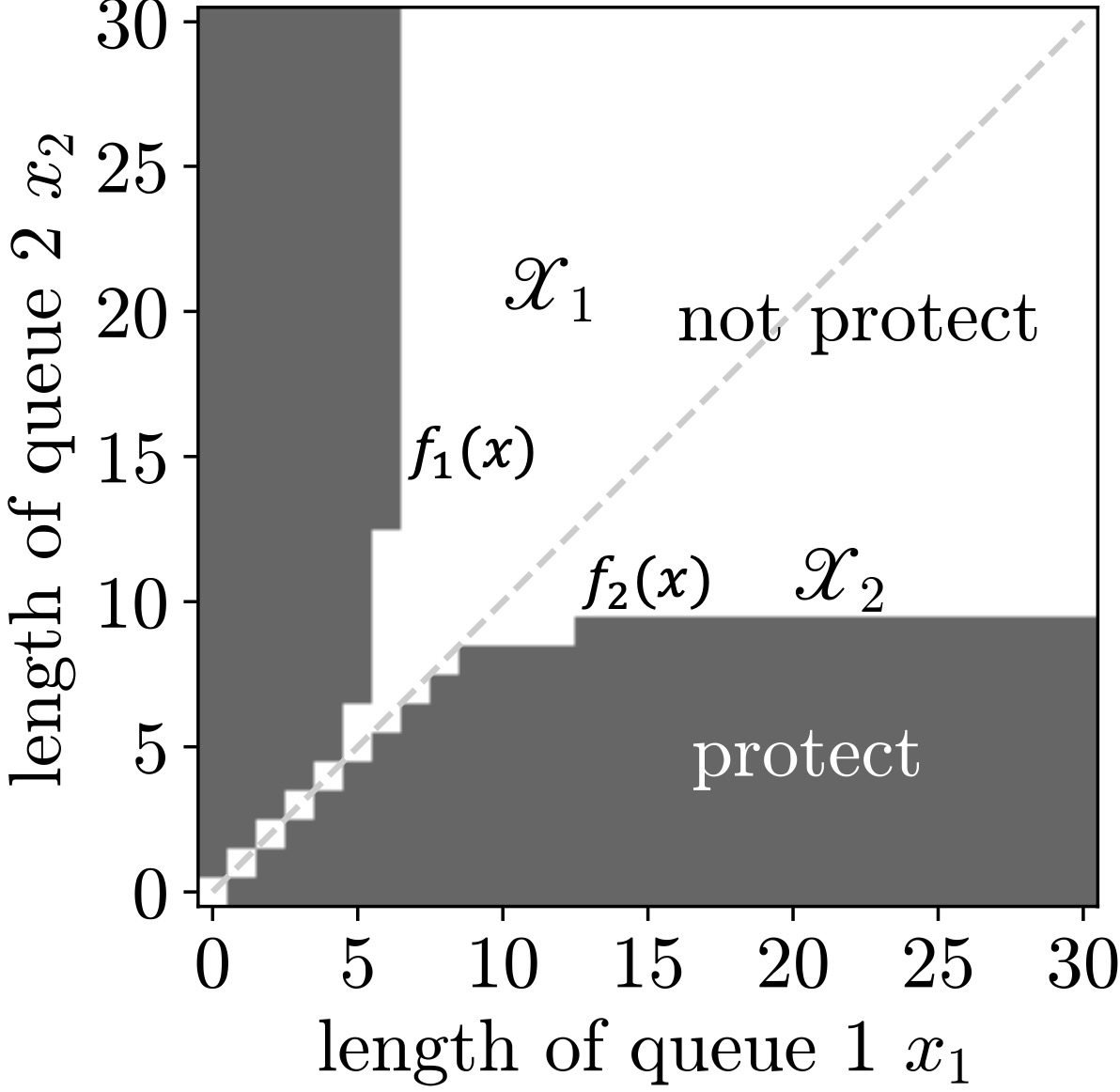}
    \caption{The characterization of the optimal protecting policy $\beta^*$ for a two-queue system ($p_1=0.1$, $p_2=0.9$, $\rho=0.5$, $a=0.9$).}
    \label{fig:protect}
\end{figure}

Here (ii) supplements (i), demonstrating that the threshold functions characterize the degree of ``imbalancedness''. They partition the state space into $n+1$ subsets: one ``inner subset'' with ``balanced'' states corresponding to action ``not protect", and the other $n$ ``outer subsets'' with ``imbalanced'' states for action ``protect". See the white and black areas in Fig.~\ref{fig:protect}. The concept ``threshold function'' has appeared in prior works \citep{bertsekas1999dynamic,hajek1984optimal,stidham1993survey}. 

The rest of this section is devoted to the proofs and discussions of Theorem~\ref{thm_stability}-\ref{thm:monotone}.

\subsection{Stability under reliability failures}\label{sec:stability}
In this subsection, we provide a proof of the stability condition under the protected case (Theorem~\ref{thm_stability}) and leave the proof of the stability condition under the unprotected case (Proposition~\ref{prp_stability}) to Appendix~\ref{sec_stability}. Both proofs use the following theorem \cite[Theorem 4.3]{meyn1993stability}:

\textbf{Foster-Lyapunov drift criterion:}
    Consider a countable-state continuous-time Markov chain $X$ with state space $S$. Let $W:S\to\mathbb{R}_{\geq0}$ be a qualified Lyapunov function, and let $\mathcal{L}$ denote the infinitesimal generator for $W$, with the drift $\mathcal{L}W$ given by
    \begin{align*}
        \mathcal{L}W(x):=\lim\limits_{t\to0}\frac{1}{t}\mathbb{E}[W(X(t))\mid X(0)=x]-W(x).
    \end{align*}
    For a non-negative function $f: S\to \mathbb{R}_{\geq0}$, if there exists $c > 0$ and $d<\infty$ and a compact set $C\subset S$ such that for every $x\notin C$, the following drift condition holds:
    \begin{align*}
     \mathcal{L}W(x)\leq-cf(x)+d,
    \end{align*}
    then for any initial condition $X(0) = x\in S$, we have
    $$\limsup\limits_{t\to\infty}\frac{1}{t}\int\limits_{\tau=0}^{t}\mathbb{E}[f(X(\tau))]d\tau\leq d/c.$$

In this paper, we care about \emph{mean boundedness}, i.e., the upper bound of the long-time average number of jobs. Thus, we consider the quadratic Lyapunov function 
\begin{align}\label{eq_quadratic}
    W(x)=\frac12\sum_{i=1}^nx_i^2,
\end{align}
and $f(x)=||x||_1$ when applying this theorem in the proofs. \\

\noindent \emph{Proof of Theorem~\ref{thm_stability}.} (i) By applying the infinitesimal generator $\mathcal{L}^{\alpha,\beta}$ to the MDP $\{X(t)\}_{t\geq0}$ under the attacking strategy $\alpha$ and the defending strategy $\beta$ as well as the Lyapunov function \eqref{eq_quadratic}, we have
\begin{align*}
\mathcal{L}^{\beta}W(x)
=& a(1-b(x))\lambda\sum\limits_{i=1}^np_i\left(W(x+e_i)-W(x)\right)\\
&+\left(1-a(1-b(x))\right)\lambda\left(W(x+e_{\min})-W(x)\right) \\
&+\mu\sum\limits_{i=1}^n\mathbbm{1}\{x_i>0\}\left(W(x-e_i)-W(x)\right) \\
=& a(1-b(x))\frac{\lambda}{2}\sum\limits_{i=1}^np_i\left((x_i+1)^2-x_i^2\right)\\
&+\left(1-a(1-b(x))\right)\frac{\lambda}{2}\left((x_{\min}+1)^2-x_{\min}^2\right) \\
&+\frac{\mu}{2}\sum\limits_{i=1}^n\mathbbm{1}\{x_i>0\}\left((x_i-1)^2-x_i^2\right) \\
= a&(1-b(x))\lambda\sum_{i=1}^n p_ix_i+\left(1-a(1-b(x))\right)\lambda x_{\min}\\
&-\mu\sum_{i=1}^n x_i+\frac12\lambda+\frac12\sum_{i=1}^n\mathbbm{1}\{x_i>0\}\mu\\
\leq a&(1-b(x))\lambda\left(\sum_{i=1}^np_ix_i-x_{\min}\right)+(\lambda x_{\min}\\
&-\mu||x||_1)+\frac12(\lambda+n\mu).
\end{align*}
By \eqref{eq_betax} there exists constants $c=\min\limits_{x\succ\boldsymbol{0}}\left\{\mu-\lambda \frac{x_{\min}}{||x||_1}-a(1-b(x))\lambda\frac{\sum_{i=1}^np_ix_i-x_{\min}}{||x||_1}\right\}>0$ and $d=\frac12(\lambda+n\mu)$ such that
\begin{align}
\mathcal{L}^{\beta}W(x)\le-c||x||_1+d,
\quad \forall x\in\mathbb Z_{\ge0}^n.
\label{eq_L^beta}
\end{align}
\footnotetext{A more rigorous definition of the infinitesimal generator $\mathcal{L}$ is a measurable function such that for each initial condition $x\in\mathbb{Z}_{\geq0}^n$ and each time $t>0$, $\mathbb{E}[W(X(t))\mid X(0)=x]=W(x)+\mathbb{E}[\int_0^t\mathcal{L}(X(\tau)d\tau\mid X(0)=x]$.}
(ii) When $\lambda<n\mu$, for every non-diagonal vector $x$, we have
$\mu||x||_1-\lambda x_{\min}>\mu||x||_1-\lambda\frac{||x||_1}{n}=\left(\mu-\frac{\lambda}{n}\right)||x||_1>0$ and
$\sum_{i=1}^np_ix_i-x_{\min}>0$,
then
$$1-\frac{\mu||x||_1-\lambda x_{\min}}{a\lambda\left(\sum\limits_{i=1}^np_ix_i-x_{\min}\right)}<1.$$
Thus, $b(x)\equiv1$ satisfies the stability condition \eqref{eq_betax} and $\beta(x)\equiv\text{P}$ is a stabilizing policy that exists.

When $\max(ap_{\max},1/n)\lambda<\mu$, for every non-diagonal vector $x$, we have
$a\lambda\sum_{i=1}^n p_ix_i+(1-a)\lambda x_{\min}\leq\max(ap_{\max},1/n)\lambda||x||_1<\mu||x||_1$,
and then
$$1-\frac{\mu||x||_1-\lambda x_{\min}}{a\lambda\left(\sum\limits_{i=1}^np_ix_i-x_{\min}\right)}<0.$$
Thus, every policy satisfies the stability criterion \eqref{eq_betax}.

By \cite[Theorem 4.3]{meyn1993stability}, this drift condition implies the upper bound \eqref{eq_Xbar_reliability} and thus the stability. 
\hfill$\square$

Theorem~\ref{thm_stability} provides a stability criterion for any state-dependent protecting probability. This implies that the operator needs to protect, i.e., choose some positive protecting probability to stabilize the system at certain states (queue lengths). We will use such stabilizing threshold probabilities to obtain a stability-constrained optimal policy. See Section~\ref{sec:constrained} and Appendix~\ref{sec:TPI}.

\subsection{Optimal protecting policy}
A standard way to solve the discounted infinite-horizon minimization problem \eqref{eq_J*} is to write down its HJB equation for optimality \cite[Chapter 4]{chang2004stochastic}:
\begin{align}
    0=\min\limits_{\beta}\{||x||_1+c_b b(x)-\gamma J^*(x)+\mathcal L^{\beta}J^*(x)\}.\label{eq:HJB}
\end{align}
We can rewrite it as the following recurrence form:
\begin{align}
    (\gamma+\lambda+n\mu)J^*(x)=&\min\limits_{\beta}\Bigg\{||x||_1+c_b b(x)+\mu\sum\limits_{i=1}^nJ^*((x-e_i)^+)+\lambda J^*(x+e_{\min})\nonumber\\
    &+(1-b(x))a\lambda\bigg(\sum\limits_{i=1}^np_iJ^*(x+e_j)-J^*(x+e_{\min})\bigg)\Bigg\}.
    \label{eq_min_beta}
\end{align}
The optimal protecting policy $\beta^*$ is essentially the solution of \eqref{eq:HJB} and \eqref{eq_min_beta}. By a standard result of discrete-state finite-action MDP \citep[Theorem 6.2.10]{puterman2014markov}, an optimal deterministic stationary policy exists. Furthermore, in the no-failure scenario ($a=0$), the operator never needs to protect (i.e., $\forall x$, $\beta^*(x)=\text{NP}$); and when all queue lengths are equal, i.e., $x_1=x_2=\cdots=x_n$, the operator deterministically deactivates the protection. \\

\noindent\emph{Proof of Theorem~\ref{thm:monotone}(i).} The expression to be minimized in the right-hand side of the HJB equation \eqref{eq_min_beta} is linear in $b(x)$, so the minimum is reached at the endpoints, that is, $b(x)=0$ or $b(x)=1$. \hfill$\square$ \\

\noindent Now the HJB equation \eqref{eq_min_beta} turns into
\begin{align*}
    (\gamma+\lambda+n\mu)J^*(x)=&\min\limits_{b\in\{0,1\}}\Big\{||x||_1+c_bb+\mu\sum\limits_{i=1}^nJ^*((x-e_i)^+)+\lambda J^*(x+e_{\min})\\
    &+(1-b)a\lambda\Big(\sum\limits_{i=1}^np_iJ^*(x+e_i)-J^*(x+e_{\min})\Big)\Big\}.
\end{align*}
Let $\tilde{\lambda}=\lambda/(\gamma+\lambda+n\mu)$, $\tilde{\mu}=\mu/(\gamma+\lambda+n\mu)$, and $\tilde{J}^*(\cdot)=(\gamma+\lambda+n\mu)J^*(\cdot)$, then we have
\begin{align}\label{eq_uniformization}
    \tilde{J}^*(x)=&\min\limits_{b\in\{0,1\}}\Big\{||x||_1+c_bb+\tilde{\mu}\sum\limits_{i=1}^n\tilde{J}^*((x-e_i)^+)+\tilde{\lambda}\tilde{J}^*(x+e_{\min}) \nonumber\\
    &+(1-b)a\tilde{\lambda}\Big(\sum\limits_{i=1}^np_i\tilde{J}^*(x+e_i)-\tilde{J}^*(x+e_{\min})\Big)\Big\}.
\end{align}



\noindent\emph{Proof of Theorem \ref{thm:monotone}(ii).}
By applying the DP recursion technique \citep[Chapter 4.6]{bertsekas1999dynamic}, we can demonstrate (a) the existence of the threshold functions by
showing (b) the monotonicity of the optimal protecting probability: $\forall x\in\mathbb Z_{\ge0}^n$, let $m=\argmin_i x_i$, then
\begin{align}\label{eq_threshold}
\begin{split}
    &b^*(x+e_i)\geq b^*(x),\quad\forall i\neq m\\
    &b^*(x-e_m)\geq b^*(x).  
\end{split}
\end{align}
Now we prove \eqref{eq_threshold}.
Let $\Delta^*(x)=\sum\limits_{i=1}^np_i\tilde{J}^*(x+e_i)-\tilde{J}^*(x+e_m)$. Note that by Definition~\ref{defn:optimal} and Theorem~\ref{thm:monotone}(i), $b^*(x)=1$ if $\Delta^*(x)>c_b/(a\tilde{\lambda})$ and $b^*(x)=0$ if $\Delta^*(x)<c_b/(a\tilde{\lambda})$.
Then the monotonicity of $b^*$ is essentially the monotonicity of $\Delta^*$. Thus, \eqref{eq_threshold} is equivalent to
\begin{align}\label{eq_Delta}
\begin{split}
    &\Delta^*(x+e_i)\geq\Delta^*(x),\quad\forall i\neq m\\
    &\Delta^*(x-e_m)\geq\Delta^*(x).
\end{split}
\end{align}
We defer the proof of \eqref{eq_Delta} to Appendix~\ref{sec:induction1}. The high-level idea is to use induction based on value iteration.
\hfill$\square$ \\

To obtain an estimated optimal policy, we propose an algorithm called truncated policy iteration (TPI). See Algorithm~\ref{alg:TPI} in Appendix~\ref{sec:TPI}. It is adapted from the classic policy iteration algorithm \citep{sutton2018reinforcement} and based on the following value iteration form of the HJB equation \eqref{eq_uniformization}:
\begin{align}\label{eq_VI}
    \tilde{J}^{k+1}(x)=&\min\limits_{b\in\{0,1\}}\Big\{||x||_1+c_bb+\tilde{\mu}\sum\limits_{i=1}^n\tilde{J}^k((x-e_i)^+)+\tilde{\lambda}\tilde{J}^k(x+e_{\min}) \nonumber\\
    &+(1-b)a\tilde{\lambda}\Big(\sum\limits_{i=1}^np_i\tilde{J}^k(x+e_i)-\tilde{J}^k(x+e_{\min})\Big)\Big\}
\end{align}
Now we can use the estimated optimal policy to conduct numerical analysis on 1) the relationship between the incentive to protect and the system parameters; 2) the comparison between the optimal policy and two naive static policies: always protect and never protects.

\begin{figure}[htbp]
    \centering
    \begin{subfigure}{0.4\textwidth}
    \includegraphics[width=\textwidth]{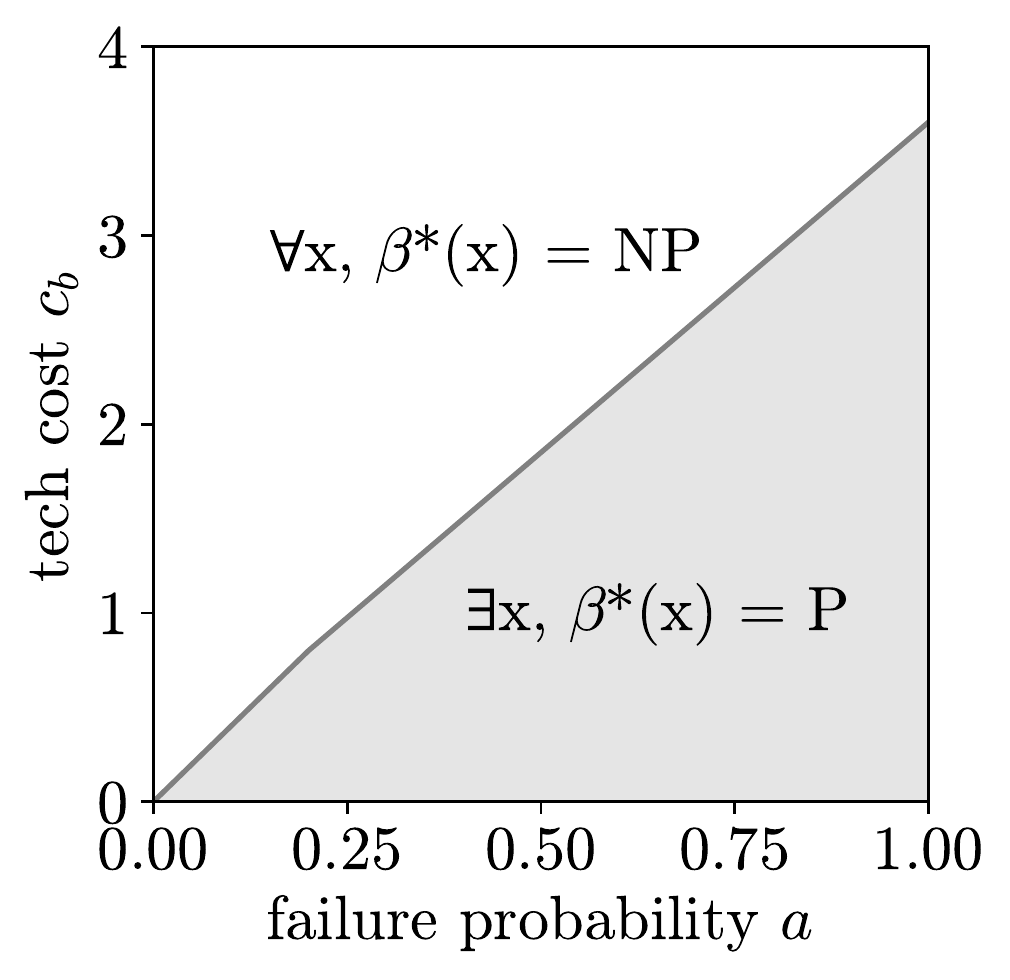}
    \subcaption{$\rho=0.8$}
    \end{subfigure}
    \begin{subfigure}{0.4\textwidth}
    \includegraphics[width=\textwidth]{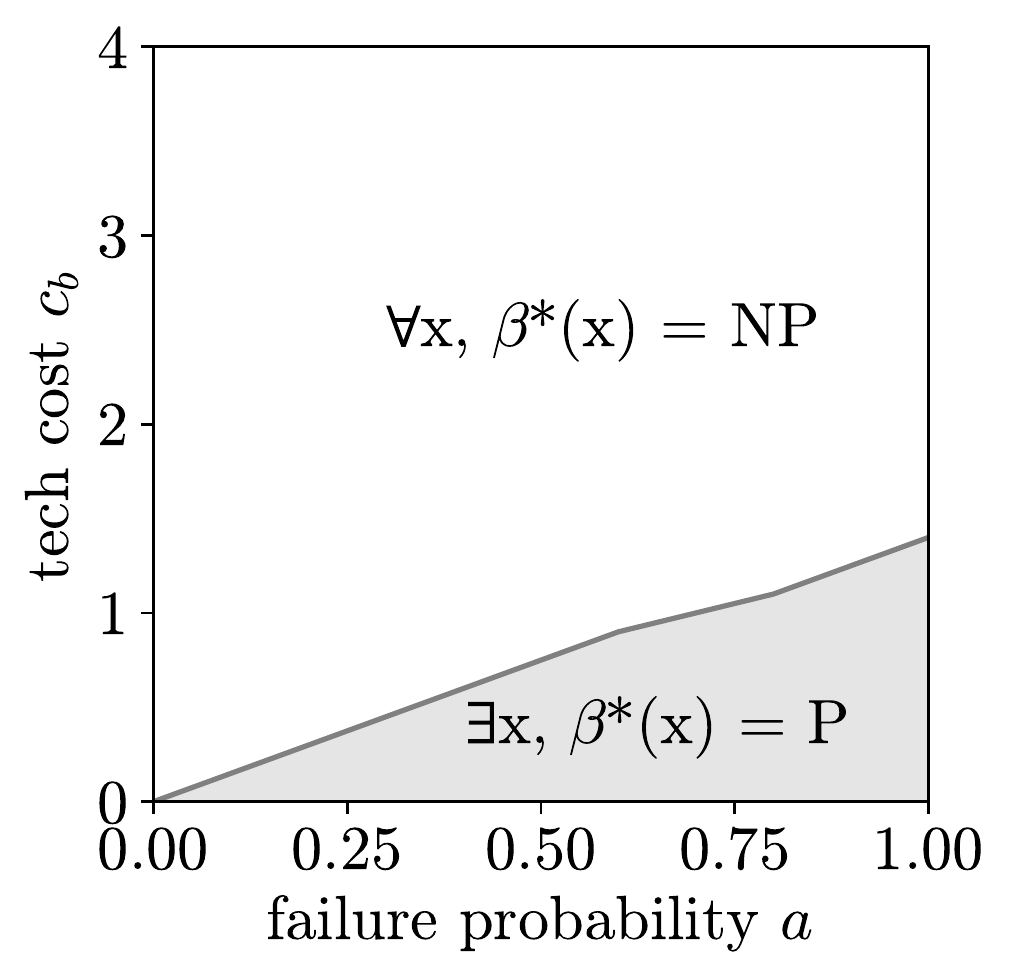}
    \subcaption{$\rho=0.2$}    
    \end{subfigure}
    \caption{The tipping points when the system operator starts to protect ``riskier'' states under the optimal policy as the failure probability and the technological cost change.}
    \label{fig:numerical}
\end{figure}

We first analyze the tipping points when the system operator starts to protect ``riskier'' states under the optimal policy $\beta^*$, i.e., $\exists x$ s.t. $\beta^*(x)=\text{P}$, as the failure probability $a$ and technological cost $c_b$ change. It can be seen from Fig.~\ref{fig:numerical} that the incentive to protect is non-decreasing in the failure probability $a$, non-increasing in the technological cost $c_b$ and non-decreasing in the \emph{utilization ratio} (a.k.a. traffic-intensity and demand-capacity ratio) $\rho=\lambda/(n\mu)$. That is, the system operator has higher incentive to protect when 1) the failure probability is higher; 2) the technological cost is lower; 3) the utilization ratio is higher.

\begin{figure}[htbp]
    \centering
    \begin{subfigure}{0.48\textwidth}
    \includegraphics[width=\textwidth]{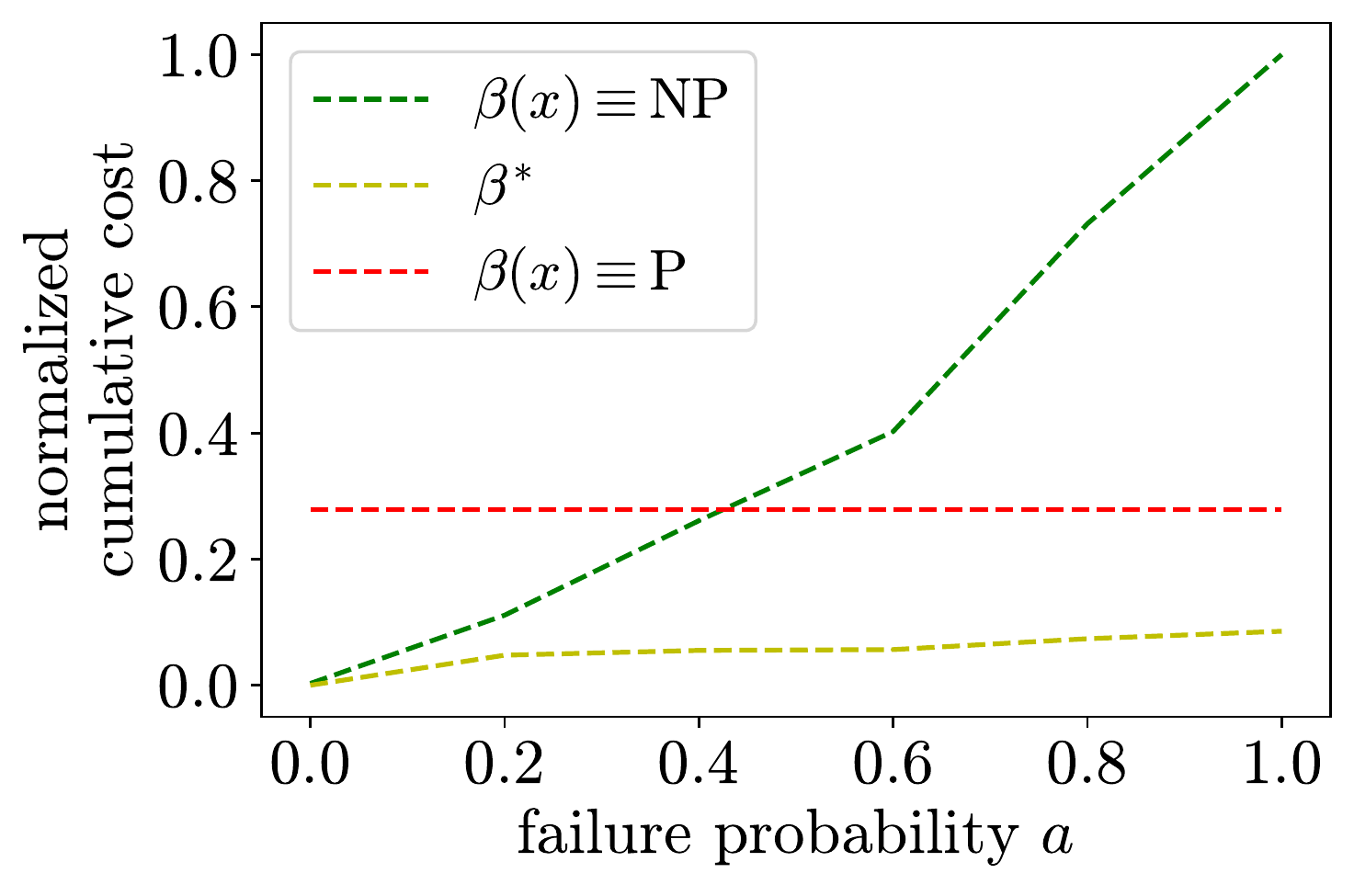}
    \subcaption{$c_b=0.05$}
    \end{subfigure}
    \begin{subfigure}{0.48\textwidth}
    \includegraphics[width=\textwidth]{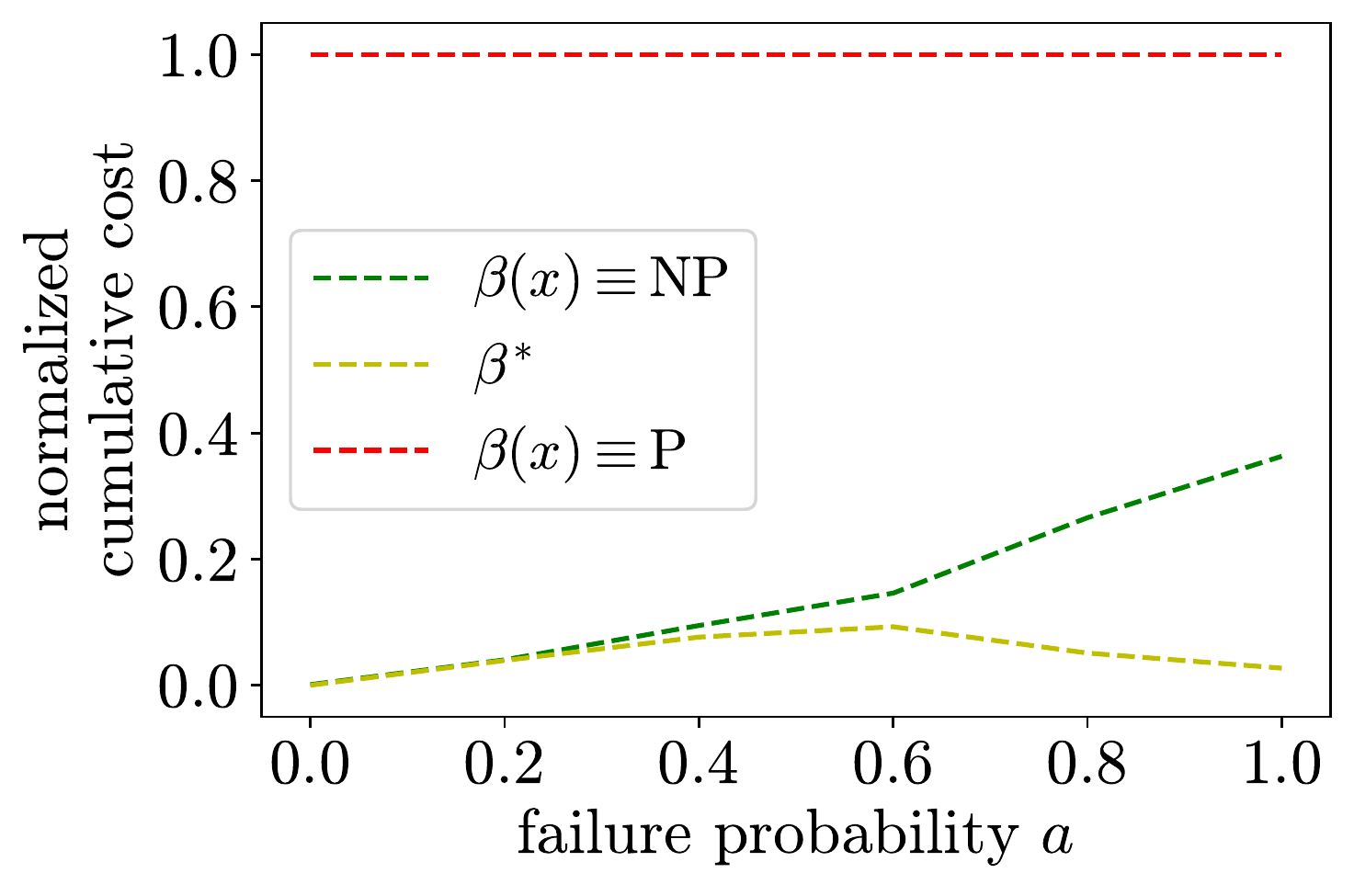}
    \subcaption{$c_b=0.5$}
    \end{subfigure}
    \caption{Comparison of the normalized cumulative discounted costs between the optimal policy and the static policies ($p_1=0.1$, $p_2=0.9$, $\rho=0.8$).}
    \label{fig:risk}
\end{figure}

In the following simulation, we will see that the optimal policy $\beta^*$ can significantly reduce the security risk, compared to the static policies: $\beta(x)\equiv\text{P}$ (always protects) and $\beta(x)\equiv\text{NP}$ (never protects). See Fig.~\ref{fig:risk} where the yellow curves are below the red curves and the green curves.
The Monte Carlo simulation result is based on the cumulative discounted cost within 50000s.
Here the cumulative discounted cost is calculated as the sum of the total queuing cost and total technological cost in the episode, and we normalized it to be a value between 0 and 1. 
Note that under the static policy $\beta(x)\equiv\text{P}$ (i.e., $\forall x$, $b(x)=1$), the job always joins the shortest queue regardless of the failure probability, so the cumulative discounted cost is a constant (red curve).

\subsection{Stability-constrained optimal policy}
\label{sec:constrained}
\begin{figure}[htbp]
    \centering
    \includegraphics[width=0.5\textwidth]{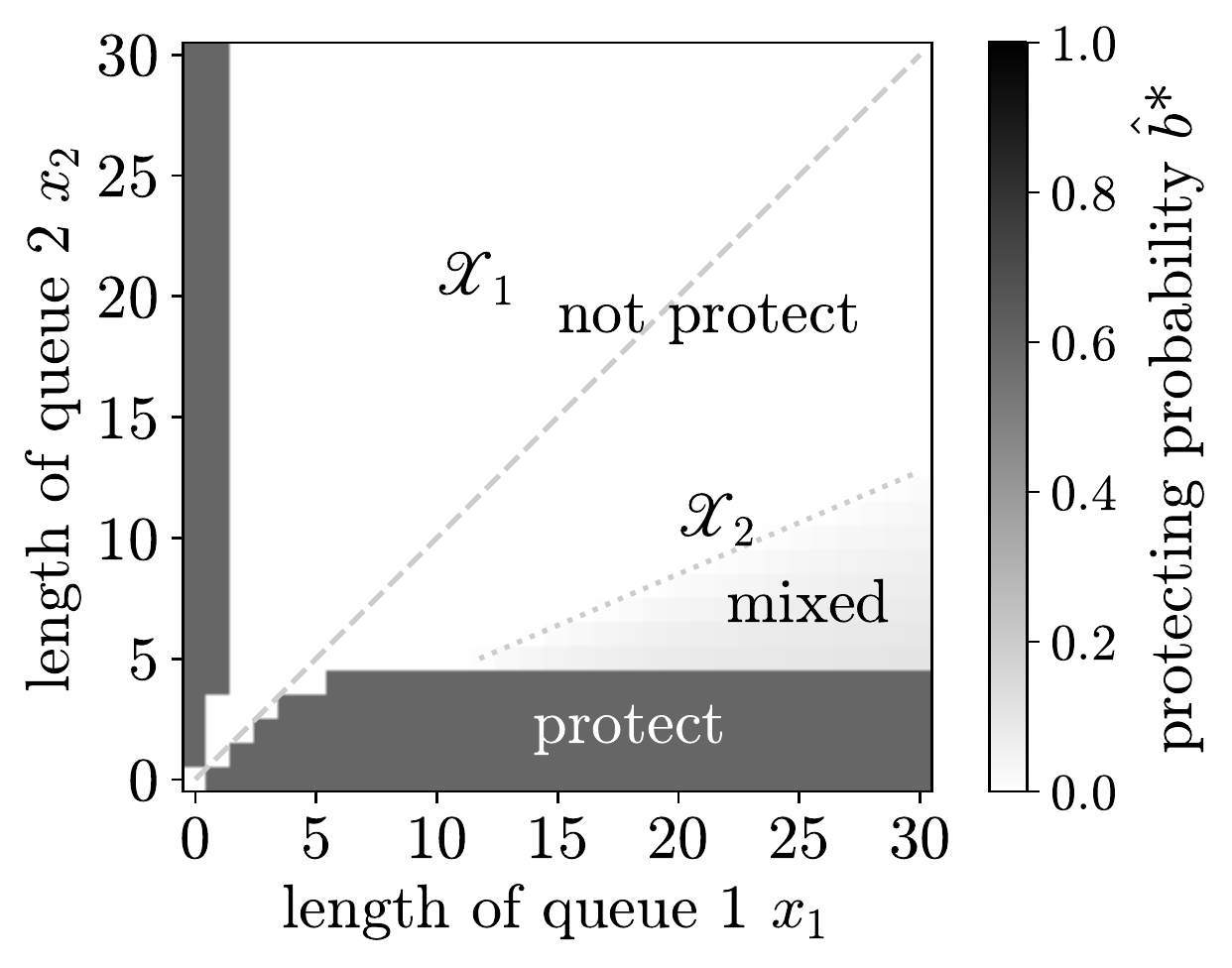}
    \caption{The characterization of the stability-constrained optimal policy $\hat{\beta}^*$ for a two-queue system ($p_1=0.1$, $p_2=0.9$, $\rho=0.5$, $a=0.9$).}
    \label{fig:stable_protect}
\end{figure}

The optimal policy may not always be stabilizing. For example, the optimal policy under the system parameters $p_1=0.1$, $p_2=0.9$, $\rho=0.5$, $a=0.9$ does not satisfy the stability condition \eqref{eq_betax}. To address this issue, we can select an optimal policy satisfying the stability condition \eqref{eq_betax} by solving a stability-constrained MDP \citep{zanon2022stability}. This involves adding an additional constraint \eqref{eq_betax} to the optimal control problem \eqref{eq_J*}. We call the solution \emph{stability-constrained optimal policy} and denote it as $\hat{\beta}^*$. This policy, unlike the optimal one, involves randomization over actions \{P, NP\} at some states, see Fig.~\ref{fig:stable_protect}. Appendix~\ref{sec:TPI} gives a corresponding modification of the TPI algorithm. 

Additionally, we can have a quick check of the existence of a stabilizing policy and stabilizability of the optimal policy using stability conditions \eqref{eq_unprotected1}-\eqref{eq_unprotected2} as follows:
\begin{itemize}
    \item When \eqref{eq_unprotected1}-\eqref{eq_unprotected2} hold, i.e., $\max(ap_{\max},1/n)\lambda<\mu$, the optimal protecting policy is also stabilizing.
    \item When only \eqref{eq_unprotected1} holds, i.e., $\lambda/n < \mu \leq ap_{\max}\lambda$, the optimal protecting policy may not be stabilizing.
    \item When \eqref{eq_unprotected1} does not hold, i.e., $\lambda \geq \mu n$, no stabilizing protecting policy exists.
\end{itemize}
\section{Defense against security failures}\label{sec:security}
In this section, we analyze the attacker's attacking strategy and system operator's defending strategy from two aspects: stability and game equilibrium.

The following criterion can be used for checking the stability of the $n$-server system under any state-dependent attacking and defending strategies:
\begin{thm}[Stability under security failures]
\label{thm:stability}
Consider an $n$-server system subject to security failures. Suppose that at each state $x\in\mathbb Z_{\ge0}^n$, the attacker (resp. system operator) attacks (resp. defends) each job following Markov strategy $\alpha$ (resp. $\beta$) characterized by a state-dependent probability $a(x)\in[0,1]$ (resp. $b(x)\in[0,1]$).
Then we have the following:
\begin{enumerate}[label=(\roman*)]
    \item The system is stable if for every non-diagonal vector $x$, the attacking and defending probabilities $a(x)$ and $b(x)$ satisfy
    \begin{align}
        a(x)\left(1-b(x)\right)<\frac{\mu||x||_1-\lambda x_{\min}}{\lambda(x_{\max}-x_{\min})}.
        \label{eq_alphax}
    \end{align}
    \item When $\lambda<n\mu$, there must exist a strategy $\beta$ with defending probability $b(x)$ satisfying \eqref{eq_alphax}.
    \item Furthermore, if \eqref{eq_alphax} holds, then the long-time average number of jobs is upper-bounded by
    \begin{align}
        \bar X\le\frac{\lambda+n\mu}{2c},
        \label{eq_Xbar_security}
    \end{align}
    where 
    $$c=\min\limits_{x\succ\boldsymbol{0}}\Big\{\mu-\lambda \frac{x_{\min}}{||x||_1}-a(x)(1-b(x))\lambda\frac{x_{\max}-x_{\min}}{||x||_1}\Big\}.$$
\end{enumerate}
\end{thm}

The next result characterizes the structure of the strategy of the Markov perfect equilibria of the stochastic game: the equilibrium defending probability is higher when the queue lengths are more ``imbalanced''.

\begin{thm}[Markov perfect equilibrium]\label{thm:equilibria}
The Markov perfect equilibrium (MPE) of the attacker-defender stochastic game exists, and the following holds:
\begin{enumerate}[label=(\roman*)]
\item $(\alpha^*,\beta^*)$ is qualitatively different over the following three subsets of the state space $\mathbb Z_{\ge0}^n$: 
\begin{enumerate}
    \item $S_1=\{x\in\mathbb Z_{\ge0}^n\mid(\alpha^*(x),\beta^*(x))=(\text{NA},\text{NP})\}$; (``low risk")
    \item $S_2=\{x\in\mathbb Z_{\ge0}^n\mid(\alpha^*(x),\beta^*(x))=(\text{A},\text{NP})\}$; (``medium risk")
    \item $S_3=\{x\in\mathbb Z_{\ge0}^n\mid(\alpha^*(x),\beta^*(x))\text{ is mixed}\}$. (``high risk")
\end{enumerate}
\item The boundaries between $S_1$ and $S_2$, as well as those between $S_2$ and $S_3$ are characterized by threshold functions $g_{ij}, h_{ij}$ ($1\leq i\neq j\leq n$) as follows:
\begin{align*}
    &S_1 = \left\{x\in\mathbb Z_{\ge0}^n \mid \bigwedge\limits_{1\leq i\neq j\leq n} (g_{ij}(x)<0)\right\}, \\
    &S_2 = \left\{x\in\mathbb Z_{\ge0}^n \mid \bigwedge\limits_{1\leq i\neq j\leq n} (g_{ij}(x)>0 \vee h_{ij}(x)<0)\right\}, \\
    &S_3 = \left\{x\in\mathbb Z_{\ge0}^n \mid \bigwedge\limits_{1\leq i\neq j\leq n} (h_{ij}(x)>0)\right\}.
\end{align*}
where for each $i, j = 1, 2, \cdots, n$ ($i\neq j$),
\begin{enumerate}
    \item $g_{ij}, h_{ij}: \mathbb Z_{\ge0}^n \to \mathbb{R}$
    separate the polyhedron  $\mathscr{X}_{ij}=\{x\in\mathbb Z_{\ge0}^n\mid x_i = x_{\max},\ x_j = x_{\min}\}$ into three subsets: $S_1\cap\mathscr{X}_{ij}$, $S_2\cap\mathscr{X}_{ij}$ and $S_3\cap\mathscr{X}_{ij}$;
    \item state $x$ has a lower (resp. higher) or equal security level than state $x+e_i$ (resp. $x+e_j$).
\end{enumerate}
\end{enumerate}
\end{thm}

\begin{figure}[H]
    \centering

    \includegraphics[width=0.48\textwidth]{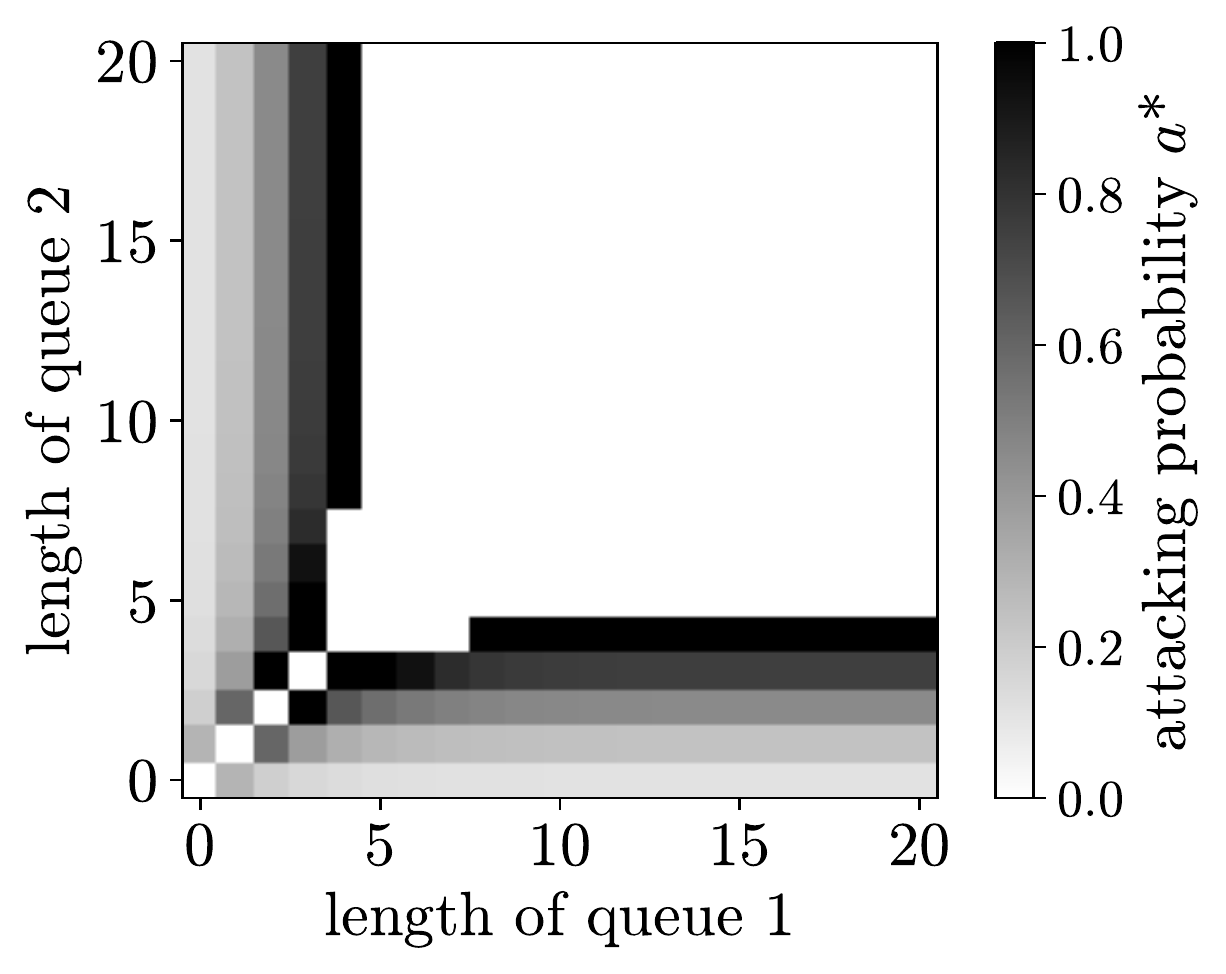}
    \includegraphics[width=0.48\textwidth]{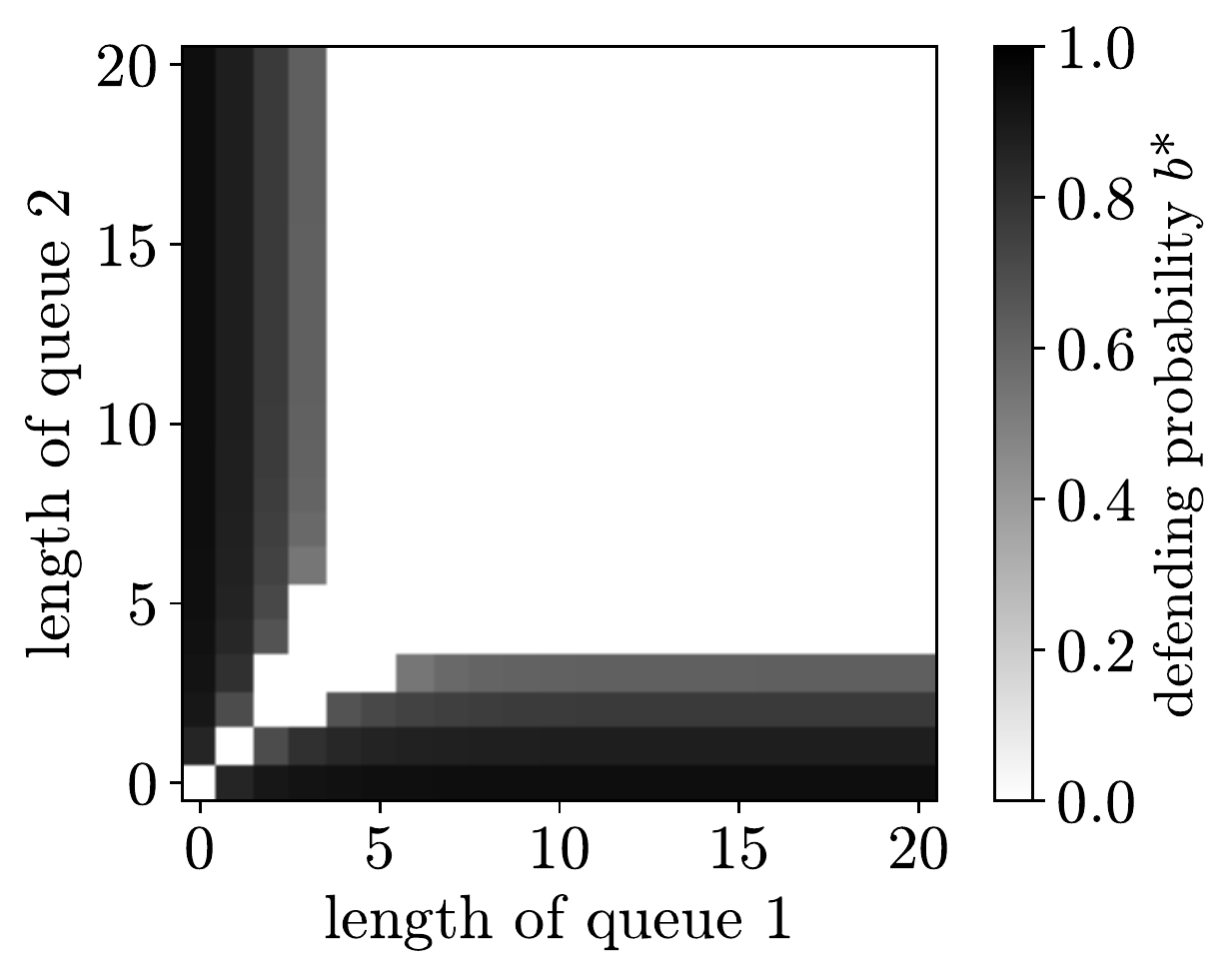}
    \caption{The equilibrium attacking and defending strategies for a two-queue system ($\rho=0.5$, $c_a=0.1$, $c_b=0.2$).}
    \label{fig:strategy}
\end{figure}

The threshold functions here also characterize the degree of ``unbalancedness''. Intuitively, $S_1$--$S_3$ correspond to various security risk levels, and thus correspond to the incentive of the operator to defend: when the queues are more ``unbalanced'', the security risk is higher, and the system operator has a higher incentive to defend.  Fig.~\ref{fig:strategy} visualizes the equilibria for a two-queue system, which help understand Theorem~\ref{thm:equilibria}. For a detailed argument about the relationship between the security levels and system parameters (e.g., technological costs and utilization ratio), see Section~\ref{sec:MPE}. 

\begin{figure}[htbp]
    \centering
    \includegraphics[width=0.95\textwidth]{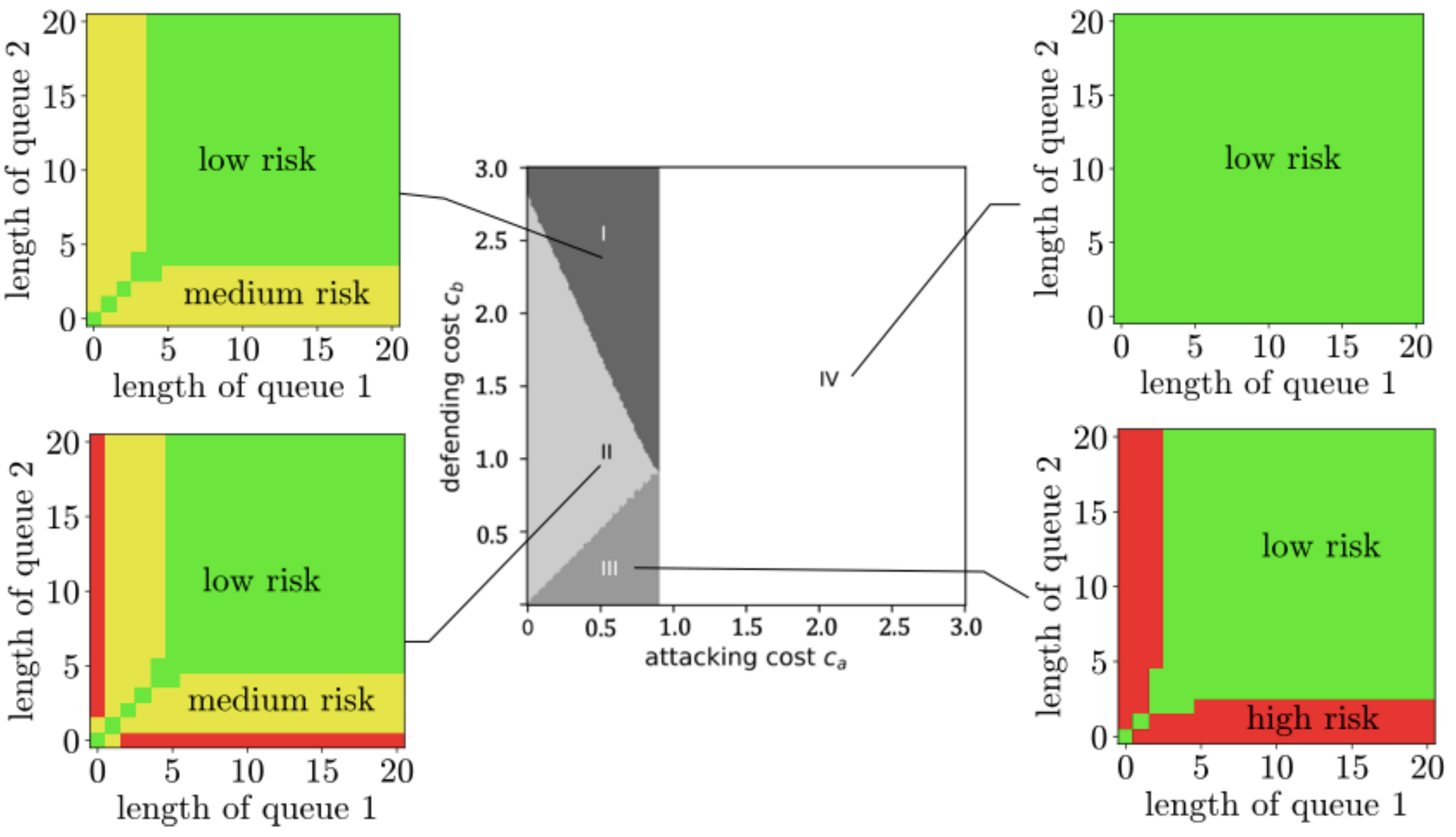}
    \caption{Equilibrium regimes of the security game ($\rho=0.8$).}
    \label{fig:equilibria regime}
\end{figure}

We also find that the security game has four equilibrium regimes under different combinations of attack cost $c_a$ and defense cost $c_b$; see Fig.~\ref{fig:equilibria regime}. Each regime is labeled with corresponding subsets of Markov perfect equilibria and security levels.

The rest of this section is devoted to the proofs of Theorem \ref{thm:stability}-\ref{thm:equilibria}, as well as an additional discussion on Markov perfect equilibrium.

\subsection{Stability under security failures}
\emph{Proof of Theorem~\ref{thm:stability}.} 
(i) By applying infinitesimal generator $\mathcal{L}^{\alpha,\beta}$ under the attacking strategy $\alpha$ and the defending strategy $\beta$ to the same Lyapunov function \eqref{eq_quadratic}, we have
\begin{align}
    \mathcal{L}^{\alpha,\beta}W(x):=&\lim\limits_{t\to0}\frac{1}{t}\mathbb{E}[W(X(t))\mid X(0)=x;\alpha,\beta]-W(x) \nonumber\\
    =&a(x)(1-b(x))\lambda\left(W(x+e_{\max})-W(x)\right)\nonumber\\
    &+\left(1-a(x)(1-b(x))\right)\lambda\left(W(x+e_{\min})-W(x)\right) \nonumber\\
    &+\mu\sum\limits_{i=1}^n\mathbbm{1}\{x_i>0\}\left(W(x-e_i)-W(x)\right) \nonumber\\
    =& a(x)(1-b(x))\frac{\lambda}{2}\left((x_{\max}+1)^2-x_i^2\right)\nonumber\\
    &+\left(1-a(x)(1-b(x))\right)\frac{\lambda}{2}\left((x_{\min}+1)^2-x_{\min}^2\right) \nonumber\\
    &+\frac{\mu}{2}\sum\limits_{i=1}^n\mathbbm{1}\{x_i>0\}\left((x_i-1)^2-x_i^2\right) \nonumber\\
    =&a(x)(1-b(x))\lambda x_{\max}+\left(1-a(x)(1-b(x))\right)\nonumber\\
    &\lambda x_{\min}-\sum_{i=1}^n\mu x_i+\frac12\lambda+\frac12\sum_{i=1}^n\mathbbm{1}\{x_i>0\}\mu\nonumber\\
    \le&a(x)(1-b(x))\lambda(x_{\max}-x_{\min})\nonumber\\
    &+\lambda x_{\min}-\mu||x||_1+\frac12(\lambda+n\mu).
    \label{eq_L^alpha,beta}
\end{align}
Hence, by \eqref{eq_alphax} there exists $c=\min\limits_{x\succ\boldsymbol{0}}\{\mu-\lambda\frac{x_{\min}}{||x||_1}-a(x)(1-b(x))\lambda\frac{x_{\max}-x_{\min}}{||x||_1})\}>0$ and $d=\frac12(\lambda+n\mu)$ such that
$$
\mathcal{L}^{\alpha,\beta}W(x)\le-c||x||_1+d,
\quad \forall x\in\mathbb Z_{\ge0}^n.
$$
(ii) When $\lambda<n\mu$, for every non-diagonal vector $x$, we have $\mu||x||_1 \geq n\mu x_{\min} > \lambda x_{\min}$, and $x_{\max}>x_{\min}$, implying $\frac{\mu||x||_1-\lambda x_{\min}}{\lambda(x_{\max}-x_{\min})}>0$. Thus, no matter what strategy the attacker chooses, the defending strategy with $b(x)\equiv1$ satisfies the stability condition \eqref{eq_alphax}. \\

\noindent (iii) By \citep[Theorem 4.3]{meyn1993stability}, the drift condition \eqref{eq_L^alpha,beta} implies the upper bound \eqref{eq_Xbar_security} and thus the stability.
\hfill$\square$

\subsection{Markov perfect equilibrium}\label{sec:MPE}
For the attacker-defender stochastic game, we first show the existence of Markov perfect equilibrium.
\begin{prop}
Markov perfect equilibrium $(\alpha^*, \beta^*)$ of the stochastic attacker-defender game always exists.
\end{prop}
\noindent\emph{Proof.} Note that the state space $\mathbb Z_{\ge0}^n$ is countable and the action space $\{0,1\}$ is finite (and thus compact). By \citep[Theorem 1]{federgruen1978n}, Markov perfect equilibrium (also called discounted equilibrium point of policies) exists. \hfill$\square$ \\

Next, we discuss the derivation of Markov perfect equilibria. According to Shapley's extension on minimax theorem for stochastic game \citep{shapley1953stochastic}, the attacker and the defender have the same equilibrium (minimax) value:
$$V_B^*(x;\alpha^*)=V_A^*(x;\beta^*)=V^*(x).$$
Thus, we only need to compute the minimax value $V^*$ of the stochastic game. Similar to the derivation of \eqref{eq_uniformization}, we obtain the following HJB equation of the minimax problem (letting $\tilde{V}^*(\cdot)=(\rho+\lambda+n\mu)V^*(\cdot)$):
\begin{align}\label{eq_Bellman_V}
    \tilde{V}^*(x)=&\max_{\alpha}\min_{\beta}\bigg\{||x||_1+c_bb(x)-c_aa(x)+\tilde{\mu}\sum\limits_{i}\tilde{V}^*((x-e_i)^+)\nonumber\\
    &+\tilde{\lambda} \tilde{V}^*(x+e_{\min})+a(x)(1-b(x))\tilde{\lambda}\Big(\tilde{V}^*(x+e_{\max})-\tilde{V}^*(x+e_{\min})\Big)\bigg\}.
\end{align}
For each state $x\in\mathbb{Z}_{\geq0}^n$, let $\delta^*(x)=\tilde{\lambda}(\tilde{V}^*(x+e_{\max})-\tilde{V}^*(x+e_{\min}))$ and build an auxiliary matrix game
\begin{align}
M(x,\tilde{V}^*)=&\Big(||x||_1+\tilde{\mu}\sum\limits_{i=1}^n\tilde{V}^*((x-e_i)^+)+\tilde{\lambda} \tilde{V}^*(x+e_{\min})\Big)
    \nonumber\\
    &\begin{bmatrix}
    1 & 1\\
    1 & 1
    \end{bmatrix}+
    \begin{bmatrix}
    0 & c_b\\
    -c_a+\delta^*(x) & -c_a+c_b
    \end{bmatrix}.
\label{eq_M}
\end{align}
Then given $\delta^*(x)$, the equilibrium strategies $(\alpha^*(x), \beta^*(x))$ can be obtained by \emph{Shapley-Snow method} \citep{shapley1952basic}, a convenient algorithm for finding the minimax value and equilibrium strategies of
any two-player zero-sum game. \\

\noindent\emph{Proof of Theorem~\ref{thm:equilibria}(i).}
Consider the matrix game $M(x, \tilde{V}^*)$ defined as \eqref{eq_M} where the attacker and the system operator are the row player and the column player. Based on Shapley-Snow method, the equilibrium strategies $(\alpha^*(x),\beta^*(x))$ are in the following three cases depending on the relationship between $\delta^*(x)$ and the technological costs $c_a,c_b>0$:

\begin{enumerate}[label=(\alph*)]
    \item When $\delta^*(x) \leq c_a$, it is obvious that $\alpha^*(x)=\text{NA}$ (i.e., $a^*(x)=0$) is a dominant strategy. Then $c_b>0$ implies $\beta^*(x)=\text{NP}$ (i.e., $b^*(x)=0$). That is, the attacker has no incentive to attack, and thus the defender does not need to defend. At this pure strategy equilibrium, the security risk is low.
    \item When the defense cost $c_b$ is higher then the attack cost $c_a$, and $c_a<\delta^*(x)\leq c_b$, it is obvious that $\beta^*(x)=\text{NP}$ (i.e., $b^*(x)=0$) is a dominant strategy. Then $c_b>-c_a+c_b$ implies $\alpha^*(x)=\text{A}$ (i.e., $a^*(x)=1$). That is, the defender has no incentive to defend and consequently the attacker prefers to attack. At this pure strategy equilibrium, the security risk is higher than the first case but tolerable.
    \item When $\delta^*(x)>\max\{c_a,c_b\}>0$, no saddle point exists. Then both the attacker and the system operator consider mixed strategies such that $a^*(x)=\frac{c_b}{\delta^*(x)}$, $b^*(x)=1-\frac{c_a}{\delta^*(x)}$. Particularly, the operator needs to select positive protecting probability, and now the security risk is high.
\end{enumerate}
The above three cases correspond to the three subsets of states. Note that the subset $S_2$ is empty when $c_a>c_b$.
\hfill$\square$ \\

From the above proof, we observe that for fixed technological costs $c_a$ and $c_b$, the security risk level is only higher when $\delta^*$ is larger. Then as in the proof of Theorem~\ref{thm:monotone}, we use the fact that the monotonicity of the security risk level is equivalent to the monotonicity of $\delta^*$ to show the threshold property of the equilibrium.
Now we are ready to present the proof of Theorem~\ref{thm:equilibria}(ii) which uses symmetry (property (i)) and Schur Convexity (property (ii)). \\

\noindent\emph{Proof of Theorem \ref{thm:equilibria}(ii).} 
For any $x\in\mathbb Z_{\ge0}^n$, let $l=\argmax_i x_i,\ m=\argmin_i x_i$. Then $\delta^*(x)=\tilde{\lambda}(\tilde{V}^*(x+e_{l})-\tilde{V}^*(x+e_{m}))$.
Since the monotonicity of the security risk level of the states is equivalent to the monotonicity of $\delta^*$, and implies the existence of the threshold functions, it is sufficient to show that $\delta^*$ is monotonically non-decreasing (resp. non-increasing) in the largest variable (resp. the smallest variable) when other variables are fixed; that is,
\begin{align}\label{eq_d}
\delta^*(x+e_{l})\geq \delta^*(x),\quad \delta^*(x-e_{m})\geq \delta^*(x). 
\end{align}
The proof of \eqref{eq_d} also uses induction based on value iteration and can be found in Appendix~\ref{sec:induction2}.
\hfill$\square$ \\

\subsection{Equilibrium computation and equilibrium regimes}
First, we discuss the numerical computation of the minimax value $V^*(x)$ and the equilibrium strategies $(\alpha^*(x), \beta^*(x))$ for each state $x$.
Based on the value iteration form of HJB equation \eqref{eq_Bellman_V}, we develop an algorithm adapted from Shapley's algorithm \citep{shapley1953stochastic,alpcan2010network,thai2016resiliency}. See Algorithm \ref{alg:Shapley} in Appendix~\ref{alg:Shapley}.

The algorithm proceeds as follows. Initialize $\tilde{V}^{0}(x)$. In each iteration $k$ and for each state $x$, let $\delta^{k}(x)=\tilde{\lambda}(\tilde{V}^{k}(x+e_{\max})-\tilde{V}^{k}(x+e_{\min}))$ and build an auxiliary matrix game $M(x, \tilde{V}^{k})$ similar to \eqref{eq_M}; then update $\tilde{V}^{k+1}(x)$ with the minimax value $val(M)$ given by Shapley-Snow method:
\begin{itemize}
    \item when $\delta^k(x)\leq c_a$, $val(M)=||x||_1+\tilde{\mu}\sum\limits_{i=1}^n\tilde{V}^k((x-e_i)^+)+\tilde{\lambda} \tilde{V}^k(x+e_{\min})$;
    \item when $c_a<\delta^k(x)\leq c_b$, $val(M)=||x||_1-c_a+\tilde{\mu}\sum\limits_{i=1}^n\tilde{V}^k((x-e_i)^+)+\tilde{\lambda} \tilde{V}^k(x+e_{\max})$;
    \item when $\delta^k(x)>\max\{c_a,c_b\}$, $val(M)=||x||_1+c_b+\tilde{\mu}\sum\limits_{i=1}^n\tilde{V}^k((x-e_i)^+)+\tilde{\lambda} \tilde{V}^k(x+e_{\min})-\frac{c_ac_b}{\delta^k(x)}$.
\end{itemize}
When $\tilde{V}^k(x)$ converges to $\tilde{V}^*(x)$, we again use Shapley-Snow method to solve the matrix game $M(x, \tilde{V}^*)$ and obtain the estimation of the equilibrium $(\alpha^*(x), \beta^*(x))$. 

Next, we discuss the existence of different security levels under different combinations of $c_a$ and $c_b$. We have seen that no medium risk state exists when $c_a>c_b$.

In Fig.~\ref{fig:equilibria regime}, various regimes correspond to particular combinations of security levels. Under large attack cost, the attacker has no incentive to attack, then only the low risk states exist (see regime IV). When the attack cost goes smaller but still greater than the defense cost ($c_a>c_b$), not only the low risk states but also the high risk states exist (see regime III) since the attacker has less incentive to attack. As the defense cost increases to be greater than the attack cost ($c_b>c_a$), the defender has less incentive to defend, and now all risk levels including the medium risk exist (see regime II).

Last, we remark that deriving a stability-constrained equilibrium is not reasonable as stability constraints can only be imposed on the system operator, not the attacker. Nevertheless, we can derive stability-constrained best responses for the operator based on \eqref{eq_alphax}, given the attacker's strategy.
\section{Concluding Remarks}\label{sec:conclude}
In this work, we address the reliability and security concerns of service systems with dynamic routing and propose cost-efficient protection/defense advice for system operators.
Moreover, our theoretical results can provide insights for real-world applications like vehicle navigation, signal-free intersection control, flight dispatch, and data packet routing. 
Interesting future directions include 1) detailed analysis of stability-constrained optimal policies/best responses; 2) extension to general queuing networks, note that stability for renewal arrival processes and general service times can be readily established using fluid model techniques \citep{dai1995stability}; 3) design of practical near-optimal heuristic policies and analysis of optimality gaps; and 4) design of efficient (in both time and space) computational algorithms that mitigate the curse of dimensionality in scenarios involving numerous parallel servers.



\section*{Acknowledgement}
The authors appreciate the discussions with Ziv Scully, Manxi Wu, Siddhartha Banerjee, Zhengyuan Zhou, Yu Tang, Haoran Su, Xi Xiong, and Nairen Cao. Undergraduate student Dorothy Ng also contributed to this project.

\bibliographystyle{apalike}        
\bibliography{bibliography}           



\appendix
\section{Appendices}
\subsection{Proof of Proposition ~\ref{prp_stability}}\label{sec_stability}
Here we provide the proof of the stability condition for unprotected system using standard results on Poisson process subdivision and the stability of generalized join the shortest queue systems \citep[Theorem 1]{foley2001join}.

\textbf{Stability of generalized JSQ:}
Let $N=\{1,2,\cdots,n\}$ be the set of $n$ exponential servers. For each nonempty subset $S\subset N$, define the traffic intensity on $S$ as
\begin{align*}
    \rho_S:=\frac{\sum_{S'\subset S}\lambda_{S'}}{\mu_S}=\frac{\sum_{S'\subset S}\lambda_{S'}}{\sum_{i\in S}\mu_i}.
\end{align*}
Let $\rho_{\max}:=\max_{S\subset N}\rho_S$ be the traffic intensity of the most heavily loaded subset. The generalized join the shortest queue system is stable if and only if $\rho_{\max}<1$.

\emph{Proof of proposition~\ref{prp_stability}}. The unprotected $n$-queue system has $n+1$ classes of jobs. 
The $i$-th class enters server $i$ as a Poisson process of rate $ap_i\lambda$ ($1\leq i\leq n$).
The $(n+1)$-th class enters the $n$-queue system as a Poisson process of rate $(1-a)\lambda$; when a job of this class arrives, the job joins the shortest queue. 
By \citep[Theorem 1]{foley2001join}, the $(n+1)$-class, $n$-queue system is stable if and only if
\begin{align*}
    \rho_S:=\frac{\sum_{S'\subset S}\lambda_{S'}}{\mu_S}=\frac{\sum_{S'\subset S}\lambda_{S'}}{\sum_{i\in S}\mu_i}.
\end{align*}
which is equivalent to \eqref{eq_unprotected1}-\eqref{eq_unprotected2}.

Consider the same quadratic Lyapunov function \eqref{eq_quadratic} and apply the infinitesimal generator, we have
\begin{align*}
\mathcal{L}W(x)
:=& \lim\limits_{t\to0}\frac{1}{t}\mathbb{E}[W(X(t))\mid X(0)=x]-W(x) \\
=& a(1-b)\lambda\sum\limits_{i=1}^np_i\left(W(x+e_i)-W(x)\right)\\
&+\left(1-a(1-b\right)\lambda\left(W(x+e_{\min})-W(x)\right) \\
&+\mu\sum\limits_{i=1}^n\mathbbm{1}\{x_i>0\}\left(W(x-e_i)-W(x)\right) \\
=& a\frac{\lambda}{2}\sum\limits_{i=1}^np_i\left((x_i+1)^2-x_i^2\right) \\
&+(1-a)\frac{\lambda}{2}\left((x_{\min}+1)^2-x_{\min}^2\right) \\
&+\frac{\mu}{2}\sum\limits_{i=1}^n\mathbbm{1}\{x_i>0\}\left((x_i-1)^2-x_i^2\right) \\
=& a\lambda\sum_{i=1}^n p_ix_i+(1-a)\lambda x_{\min}-\mu\sum_{i=1}^n x_i+\frac{\lambda}{2} +\frac{\mu}{2}\sum_{i=1}^n\mathbbm{1}\{x_i>0\} \\
\le&\left(\max(ap_{\max},1/n)\lambda-\mu\right) ||x||_1+\frac12(\lambda+n\mu).
\end{align*}
Hence, by \eqref{eq_unprotected1}--\eqref{eq_unprotected2} there exists a constant $c=\mu-\max(ap_{\max},1/n)\lambda>0$ and $d=\frac12(\lambda+n\mu)$ such that
\begin{align*}
    \mathcal{L}W(x)\le-c||x||_1+d,
    \quad \forall x\in\mathbb Z_{\ge0}^n.
\end{align*}
By \citep[Theorem 4.3]{meyn1993stability}, this drift condition implies the upper bound
and thus the stability. \hfill$\square$

\subsection{Induction part of Theorem~\ref{thm:monotone}}
\label{sec:induction1}
In this subsection, we continue the proof of Theorem~\ref{thm:monotone} with the induction proving \eqref{eq_Delta}.

Let $\Delta^k(x)=\sum\limits_{i=1}^np_i\tilde{J}^k(x+e_i)-\tilde{J}^k(x+e_m)$, it is sufficient to show for all $k\in\mathbb{N}$,
\begin{align*}
\begin{split}
  \Delta^k(x+e_i)&\geq\Delta^k(x),\quad\forall i\neq m\\
    \Delta^k(x-e_m)&\geq\Delta^k(x).  
\end{split}
\end{align*}
One can verify that the above hold for $k=0,1,2$ with the fact that $\tilde{J}^0=0$, $\tilde{J}^1(x)=||x||_1$ and $\tilde{J}^2(x)=(1++n\mu)||x||_1+-\mu\sum_i\mathbbm{1}\{x_i>0\}$. Here we consider multiple base cases to avoid reaching trivial conclusions, say all inequalities are just equalities.

Now we show the inductive step. According to the value iteration \eqref{eq_VI}, we have $\forall j\neq m$,
\begin{align}
    &\Delta^{k+1}(x+e_j)-\Delta^{k+1}(x)\nonumber\\
    =&\mu\sum_{i=1}^n[\Delta^k((x+e_j-e_i)^+)-\Delta^k((x-e_i)^+)]\label{eq_inductive1}\\
    &+[\Delta^k(x+e_j+e_m)-\Delta^k(x+e_m)]\label{eq_inductive2}\\
    &+f^k(x+e_j)-f^k(x),\label{eq_inductive3}\\
    &\nonumber\\
    &\Delta^{k+1}(x-e_m)-\Delta^{k+1}(x)\nonumber\\
    =&\tilde{\mu}\sum_{i=1}^n[\Delta^k((x-e_m-e_i)^+)-\Delta^k((x-e_i)^+)]\label{eq_inductive4}\\
    &+[\Delta^k(x)-\Delta^k(x+e_m)]\label{eq_inductive5}\\
    &+f^k(x-e_m)-f^k(x),\label{eq_inductive6}
\end{align}
where 
$f^k(x):=\sum_{i=1}^np_i\min\Big\{c_b,a\Delta^k(x+e_i)\Big\}-\min\Big\{c_b,a\Delta^k(x+e_m)\Big\}$.

Note that based on the induction hypothesis, we have $\forall j\neq m$,
\begin{align}
    \Delta^k((x+e_j-e_i)^+)&\geq\Delta^k((x-e_i)^+), \label{eq_hypo1}\\
    \Delta^k(x+e_j+e_m)&\geq\Delta^k(x+e_m), \label{eq_hypo2}\\
    \Delta^k(x+e_j+e_i)&\geq\Delta^k(x+e_i), \label{eq_hypo3}\\
    \Delta^k((x-e_m-e_i)^+)&\geq\Delta^k((x-e_i)^+), \label{eq_hypo4}\\
    \Delta^k(x)&\geq\Delta^k(x+e_m), \label{eq_hypo5}\\
    \Delta^k(x-e_m+e_i)&\geq\Delta^k(x+e_i).\label{eq_hypo6}
\end{align}
Then \eqref{eq_hypo1} and \eqref{eq_hypo4} naturally lead to $\eqref{eq_inductive1}\geq0$ and $\eqref{eq_inductive4}\geq0$ respectively. We can also use \eqref{eq_hypo2}-\eqref{eq_hypo3} to discuss the possibilities of \eqref{eq_inductive3} under the min operations and derive $\eqref{eq_inductive2}+\eqref{eq_inductive3}\geq0$. For example, when $\Delta^k(x+e_j+e_i)\geq\Delta^k(x+e_i)\geq\frac{c_b}{a}$ and $\Delta^k(x+e_j+e_m)\geq\Delta^k(x+e_m)\geq\frac{c_b}{a}$, we have $\eqref{eq_inductive2}+\eqref{eq_inductive3}=a\sum_{i=1}^np_i\left(\Delta^k(x+e_j+e_i)-\Delta^k(x+e_i)\right)+(1-a)\left(\Delta^k(x+e_j+e_m)-\Delta^k(x+e_m)\right)\geq0$. Similarly, we can obtain $\eqref{eq_inductive5}+\eqref{eq_inductive6}\geq0$ using \eqref{eq_hypo5}-\eqref{eq_hypo6}.

Thus, we can conclude that $\eqref{eq_inductive1}+\eqref{eq_inductive2}+\eqref{eq_inductive3}\geq0$ and $\eqref{eq_inductive4}+\eqref{eq_inductive5}+\eqref{eq_inductive6}\geq0$, which yield
$$\Delta^{k+1}(x+e_j)\geq\Delta^{k+1}(x),\quad \forall j\neq m,$$
$$\Delta^{k+1}(x-e_m)\geq\Delta^{k+1}(x).$$

\subsection{Induction part of Theorem~\ref{thm:equilibria}}
\label{sec:induction2}
In this subsection, we continue the proof of Theorem~\ref{thm:equilibria} with the induction proving \eqref{eq_d}.

Let $\delta^k(x)=\tilde{\lambda}(\tilde{V}^k(x+e_{l})-\tilde{V}^k(x+e_{m}))$, it is sufficient to show for all $k\in\mathbb{N}$,
\begin{align*}
\delta^k(x+e_{l})\geq \delta^k(x),\quad \delta^k(x-e_{m})\geq \delta^k(x). 
\end{align*}
For the base cases, one can verify that the above inequalities hold for $k=0,1,2$.
Now we show the inductive step.
According to the value iteration form of formula \eqref{eq_Bellman_V},
\begin{align*}
    &\delta^{k+1}(x+e_{l})-\delta^{k+1}(x)\nonumber\\
    =&\tilde{\mu}[\delta^k(x)-\delta^k((x-e_{l})^+)]\nonumber\\
    +&\tilde{\mu}[\delta^k((x+e_{l}-e_{m})^+)-\delta^k((x-e_{m})^+)]\\
    +&\tilde{\lambda}[\delta^k(x+e_{l}+e_{m})-\delta^k(x+e_{m})]\\
    +&g^k(x+2e_{l})-g^k(x+e_{l}+e_{m})-g^k(x+e_{l})+g^k(x+e_{m}),
\end{align*}
where 

\[g^k(x)=\max\Big\{0,\min\Big\{\delta^k(x)-c_a,c_b-\frac{c_ac_b}{\delta^k(x)}\Big\}\Big\}.\]
Note that based on the induction hypothesis, we have

$$\delta^k((x+e_{l}-e_{m})^+)\geq \delta^k((x-e_{m})^+)\geq \delta^k(x)\geq \delta^k((x-e_{l})^+),$$
$$\delta^k(x+2e_{l})\geq \delta^k(x+e_{l})\geq \delta^k(x+e_{l}+e_{m})\geq \delta^k(x+e_{m}).$$
\\
Then we can conclude that $\delta^{k+1}(x+e_{l})\geq \delta^{k+1}(x)$
and prove $\delta^{k+1}(x-e_{m})\geq\delta^{k+1}(x)$ in a similar way.

\subsection{Truncated policy iteration}\label{sec:TPI}
In this subsection, we present the truncated policy iteration algorithm for estimating stability-constrained optimal policy. This algorithm is adapted from the classic policy iteration algorithm \citep{sutton2018reinforcement} by combining the stability condition \eqref{eq_betax}. Since the original state space is countably infinite, here we set a boundary to make the state space finite so that the algorithm can terminate in finite steps.
\begin{algorithm}[H]
\caption{Truncated policy iteration for estimating optimal policy $\beta\approx\beta^*$ (continuing)}
Algorithm parameter: small $\epsilon > 0$\;\\
Input: arrival rate $\lambda$, service rate $\mu$, discounted factor $\rho$, number of queues $n$, unit protection cost $c_b$, probabilities of joining queues $p_1, p_2, \cdots, p_n$\;\\
Initialize arrays: $J(x)\in\mathbb{R}$ and $b(x)\in\{0,1\}$ arbitrarily (e.g., $J(x)=0$, $b(x)=0$) for all $x \in \mathcal{X}=\{0,1,\cdots,B\}^n$ \quad \# $B$ is the queue length upper bound \footnotemark[1]

$\tilde{\lambda}\leftarrow\lambda/(\rho+\lambda+n\mu),\quad \tilde{\mu}\leftarrow\mu/(\rho+\lambda+n\mu)$ \\
    \Repeat{$stable=True$}{
        \Repeat{$|v-V(x)|<\epsilon$}{
            $\Delta\leftarrow0$\quad
            \# Policy evaluation\\
            \ForEach{$x\in\mathcal{X}$}{
                $v\leftarrow J(x)$\\
                $J(x)\leftarrow \min\limits_{b\in\{0,1\}}\bigg\{||x||_1+c_bb+\tilde{\mu}\sum\limits_{i=1}^nJ((x-e_i)^+)+\tilde{\lambda} J(x+e_{\min}) +(1-b)a\tilde{\lambda}\bigg[\sum\limits_{i=1}^np_iJ(x+e_i)-J(x+e_{\min})\bigg]\bigg\}$ \\
                $\Delta\leftarrow\max(\Delta,|v-J(x)|)$
            }
        }
    
        $stable\leftarrow True$\quad \# Policy improvement\\
        \ForEach{$x\in\mathcal{X}$}{
            old-action$\leftarrow b(x)$\\
            \If{$a\tilde{\lambda}\left(\sum\limits_{i}p_iJ(x+e_i)-J(x+e_{\min})\right)\leq c_b$}{
                $b(x)=0$ \footnotemark[2]
            }
            \Else{
                $b(x)=1$
            }
            \If{old-action$\neq b(x)$}{
                $stable\leftarrow False$
            }
        }
    }
    Output $\beta=(1-b)\text{NP}+b\text{P}\approx\beta^*$
\label{alg:TPI}
\end{algorithm}
\footnotetext[1]{Since the estimation errors are relative large at the boundary, we can set $B$ to be larger than the real upper bound.}
\footnotetext[2]{For computing the stability-constrained optimal policy, we can modify the algorithm as follows. In the initialization, for every $x\in\mathcal{X}$, if $x_1=\cdots=x_n$ then continue, else set $\theta(x) = \max\Bigg(1-\frac{\tilde{\mu}||x||_1-\tilde{\lambda} x_{\min}}{a\tilde{\lambda}\Big(\sum\limits_{i=1}^np_ix_i-x_{\min}\Big)}, 0\Bigg)$. Then in the policy improvement, replace $b(x)=0$ with $b(x)=\theta(x)$.}

\subsection{Adapted Shapley's algorithm}
Here we present the adapted Shapley's algorithm for computing the minimax value and equilibrium strategies of the attacker-defender stochastic game. In each iteration, it builds an auxiliary matrix game and obtains the minimax value using Shapley-Snow method \citep{shapley1952basic}.
\begin{algorithm}[H]
Set $V(x)=0$ for all $x\in \mathcal{X}=\{0,1,2,\cdots,B\}^n$ \footnotemark[1]\\
\Repeat{$|v-V(x)|<\epsilon$}{
    \ForEach{$x\in\mathcal X$}{
        $v\leftarrow V(x)$\\
        Build auxiliary matrix game $M(x,V)$\\
        Compute the minimax value $val(M)$ by using \emph{Shapley-Snow method}\\
        $V(x)\leftarrow val(M)$
    }
}
\ForEach{$x\in\mathcal X$}{
    Build auxiliary matrix game $M(x,V)$\\
    Compute $\alpha$ and $\beta$ from $M(x,V)$ by using \emph{Shapley-Snow method}
}
\caption{Adapted Shapley's algorithm for estimating equilibrium strategies $\beta\approx\beta^*$, $\alpha\approx\alpha^*$ (continuing)}
\label{alg:Shapley}
\end{algorithm}
\end{document}